\documentclass[aps,preprint,showkeys]{revtex4}

\usepackage{amsmath}
\usepackage{amsfonts}
\usepackage{amssymb}
\usepackage{amsthm}
\usepackage{mathrsfs}
\usepackage{graphicx}
\usepackage{color}
\usepackage{physics}
\usepackage{float}
\usepackage{booktabs}

\usepackage[colorlinks=true,linkcolor=blue,citecolor=blue]{hyperref}

\usepackage{placeins}

\begin{document}

\title{Thin accretion-disc transfer functions in the q-metric spacetime: quadrupolar lensing signatures}

\author{Saken Toktarbay$^{1}$, Manas Khassanov $^{2}$, Aray Muratkhan$^{1}$, Zhantay Muratkhan $^{1,2}$, Aliya Taukenova$^{1*}$, Hernando Quevedo$^{3,4}$}

\affiliation{$^1$ Department of Theoretical and Nuclear Physics, Al-Farabi Kazakh National University, Almaty 050040, Kazakhstan \\
$^2$ Department of Physics, Kazakh National Women's Teacher Training University, Almaty, 050000, Kazakhstan \\
$^3$ Instituto de Ciencias Nucleares, Universidad Nacional Aut\'onoma de M\'exico, AP 70543, M\'exico, DF 04510, Mexico \\
$^4$Dipartimento di Fisica and ICRA, Universit\`a di Roma ``La Sapienza", I-00185 Roma, Italy \label{addr4}\\
{$*$}: alyiya.Taukenova@kaznu.edu.kz
}

\begin{abstract}
We study geometrically thin equatorial disc images in the static,
axisymmetric q-metric spacetime using backward ray tracing.
The Schwarzschild solution is recovered at $q=0$; for the nonzero
deformations considered here, $r=2M$ is a curvature singularity
rather than an event horizon.
For a distant nearly face-on observer, we construct the first three
equatorial transfer functions $r_m(b;q)$. In the {Schwarzschild} limit,
the resolved third-crossing interval contains the critical impact
parameter $b=3\sqrt{3}M$. As $q$ increases from $-0.3$ to $0.3$,
the branch center shifts from $b_{3,c}/M\simeq3.716$ to $6.702$.
In units of the physical monopole mass $\mathcal M=(1+q)M$,
it instead decreases from $b_{3,c}/\mathcal M\simeq5.309$ to $5.156$.
Much of the displacement in metric-parameter units therefore
reflects the changing monopole scale.
At fixed $\mathcal M$, the inclined 
first-intersection maps  show that
the contour dimensions increase with $q$, with larger relative
changes near the inner disc. A prescribed power-law emissivity
and relativistic frequency-shift weighting yield direct images
whose integrated flux differs from the Schwarzschild value by
at most $2.4\%$. Between the endpoint models, the horizontal
median-flux position increases by $8.8\%$ and the interquartile
width by $14.5\%$. Within this emission prescription, the residual
deformation dependence appears mainly as a displacement and
broadening of the brightness distribution. Higher-order radiative
contributions are not included.
\end{abstract}

\keywords{q- metric, thin equatorial disc, transfer function, Radiative maps}
\maketitle


\section{Introduction}
\label{sec:introduction}

Millimetre very-long-baseline interferometry has made horizon-scale
strong-field lensing accessible to observation. The Event Horizon Telescope
images of M87* and Sgr~A* show compact emission regions with central
brightness depressions and ring-like structures
\citep{Akiyama2019I,Akiyama2019VI,Akiyama2022I}. These structures cannot be
identified directly with properties of a spacetime metric, because their
morphology depends jointly on null-geodesic propagation, the distribution
and motion of the emitting plasma, optical-depth effects, gravitational and
Doppler frequency shifts, and the angular response of the observing array.
The distinction between the geometrical lensing map and the radiative
weighting of that map has been present since the early calculations of
relativistic disc images and black-hole shadows
\citep{Falcke2000,Luminet1979}, and it remains relevant to current studies of
photon-ring substructure \citep{Gralla2019,Johnson2020}.

A geometrically thin equatorial disc provides a controlled setting in which
the propagation and emission contributions can be analysed separately. The
geometrical part of the image is described by transfer functions that relate
a coordinate on the observer screen to the radius at which a
backward-traced photon intersects the disc
\citep{Cunningham1975,Luminet1979}. For a nearly face-on observer, the screen
coordinate can be represented by an impact parameter $b$, and successive
equatorial intersections define the ordered functions
\[
r_m(b), \qquad m=1,2,3,\ldots .
\]
The first branch describes the direct disc image, the second gives a
lensed-ring contribution, and the third is the first higher-order branch
formed by trajectories propagating close to the critical lensing region.
The corresponding observed intensity is obtained only after an emissivity
profile, an emitting domain, and the emitter-to-observer frequency shift are
specified. The transfer functions therefore determine the disc-to-screen
mapping, while the radiative prescription determines how this mapping is
weighted in an image.

The spacetime considered in this work is the Zipoy--Voorhees solution,
referred to here as the $q$-metric. It is an exact, static, axially
symmetric, asymptotically flat vacuum spacetime. The Schwarzschild solution
is recovered when the deformation parameter vanishes, whereas for nonzero
deformation the spacetime possesses an independent mass quadrupole
\citep{Zipoy1966,Voorhees1970,Quevedo1989,Quevedo2011,ToktarbayQuevedo2014}.
The same geometry, and closely related parametrizations such as the
$\gamma$-metric, have been used to analyse relativistic multipole moments,
null and timelike geodesics, circular-orbit stability, compact-object
mimickers, shadows, and accretion-disc observables
\citep{Boshkayev2016,FrutosAlfaro2018,ArrietaVillamizar2021,Boshkayev2021,
Toktarbay2022EPJC,Shaikh2022,Turimov2023,Faraji2025,Momynov2025}.
Zipoy--Voorhees-type geometries have also been considered in dynamical and
finite-distance imaging comparisons
\citep{Destounis2023,Toktarbay2026PRD}.

For $q\neq0$, the surface $r=2M$ is a curvature singularity rather than
a regular event horizon. The $q$-metric is therefore not used here as a
complete black-hole spacetime, but as an exact exterior vacuum geometry with
a controllable static quadrupolar deformation. This restriction is relevant
to the terminology adopted below. The central dark region in an image is
described as an apparent dark region, and the higher-order
disc-intersection structure is referred to as photon-ring-like, rather than
as a black-hole shadow or photon ring in the strict sense.

Previous analyses of the $\gamma$-metric have shown that singular
quadrupolar exteriors can produce shadow-like regions and thin-disc images
whose structure depends on the existence and properties of unstable null
orbits \citep{Shaikh2022}. Disc images in naked-singularity and other
compact-object-mimicker spacetimes can likewise differ from their
Schwarzschild and Kerr counterparts when the near-centre geodesic structure
is modified \citep{ShaikhJoshi2019,Gyulchev2020,Muratkhan2022}. The present
work addresses a more restricted question. We analyse how the first three
equatorial disc-crossing branches change with the quadrupole parameter and
compare the resolved higher-order branch with the analytic equatorial photon
critical scale. The novelty is therefore not the construction of a
thin-disc image in the $q$-metric itself, but the branch-resolved
transfer-function analysis, its mass-normalized interpretation, and its
connection with inclined screen maps and a minimal radiative realization.

A central issue in this comparison is the distinction between the metric
parameter $M$ and the physical monopole mass. In the convention used here,
\[
\mathcal{M}=M(1+q).
\]
Consequently, a displacement measured in units of $M$ contains both a
change of the physical monopole scale and a residual effect associated with
the quadrupolar geometry. We therefore report the relevant
impact-parameter scales in both $M$ and $\mathcal M$ units. The
fixed-$\mathcal M$ comparison is used to identify the part of the
disc-to-screen mapping that remains after the monopole rescaling has been
removed.

The calculation is organized in three connected stages. First, we construct
the first three transfer functions for a distant nearly face-on observer,
validate the Schwarzschild limit, and compare the centre of the resolved
third branch with the analytic photon critical impact parameter. Second, we
compute inclined first-intersection maps at fixed physical monopole mass and
extract iso-radial contour diagnostics. These maps probe the direct
two-dimensional disc-to-screen mapping and are not identified with the
higher-order third branch. Third, we supplement the geometrical analysis
with a minimal direct-image calculation based on a prescribed emissivity and
the relativistic frequency-shift factor. This radiative realization is not
intended as a self-consistent plasma model; its purpose is to show how the
underlying geometrical differences enter a weighted thin-disc image.

The numerical results show that the centre of the third disc-intersection
branch shifts monotonically towards larger $b/M$ as $q$ increases and
follows the deformation dependence of the analytic photon critical scale.
The displacement is substantially reduced when expressed in units of the
fixed physical monopole mass $\mathcal M$, demonstrating that most of the
variation in metric-parameter units is associated with monopole rescaling.
A residual quadrupole dependence remains in the higher-order branch and in
the inclined disc-to-screen contours, with the largest contour response
occurring for the smallest mapped disc radii. The direct radiative maps
provide a controlled illustration of how these residual geometrical changes
are modified by emissivity and frequency-shift weighting.

This paper is organized as follows. Section~\ref{sec:qmetric} introduces the
$q$-metric convention, its monopole and quadrupole moments, and the
equatorial photon and circular-orbit scales.
Section~\ref{sec:raytracing} describes the Hamiltonian ray-tracing scheme,
the observer screen, and the disc-crossing algorithm.
Section~\ref{sec:transfer} defines the transfer functions and presents the
Schwarzschild validation criterion. The face-on transfer-function results
are given in Sec.~\ref{sec:tf_results}. Section~\ref{sec:diagnostics} presents the inclined first-intersection maps,
the contour diagnostics, and the direct-image comparison.
Section~\ref{sec:emission} introduces the emissivity and frequency-shift
prescription used for the minimal radiative realization.
The physical interpretation and limitations are discussed in Sec.~\ref{sec:discussion},
and the main conclusions are summarized in Sec.~\ref{sec:conclusions}.

\section{The q-metric and the relevant strong-field scales}
\label{sec:qmetric}

We adopt the Zipoy--Voorhees spacetime, commonly referred to as the
$q$-metric, as a static axisymmetric geometry with an independent
quadrupolar deformation. In the convention used throughout this work, the
metric is

\begin{align}   \nonumber
ds^2
& =
F^{1+q}dt^2
-
F^{-q}
\bigg[
A^{-q(2+q)}
\left(
\frac{dr^2}{F}+r^2d\theta^2 \right)
\\
&
\qquad\qquad
+
r^2\sin^2\theta\,d\phi^2
\bigg],
\label{eq:q_metric}
\end{align}
where
\begin{equation}
F(r)=1-\frac{2M}{r},
\qquad
A(r,\theta)=1+\frac{M^2\sin^2\theta}{r^2-2Mr}.
\label{eq:metric_functions}
\end{equation}
Geometrized units ($G=c=1$) and the metric signature $+---$ are used
throughout. The Schwarzschild solution is recovered for $q=0$, whereas for
$q\neq0$ the surface $r=2M$ becomes a curvature singularity instead of a
regular event horizon. In the present work the $q$-metric is therefore used
as an exact vacuum exterior with an adjustable quadrupole moment. This is
sufficient for our purposes because the transfer functions are determined
entirely by null-geodesic propagation outside the singular surface and by
their intersections with a geometrically thin equatorial disc.

The parameter $M$ appearing in Eq.~\eqref{eq:q_metric} represents the metric
length scale rather than the physical monopole mass. The first two non-vanishing
relativistic multipole moments are

\begin{equation}
\mathcal{M}=M(1+q),
\label{eq:monopole_mass}
\end{equation}
and
\begin{equation}
\mathcal{Q}
=
-\frac{1}{3}M^3q(1+q)(2+q).
\label{eq:quadrupole_moment}
\end{equation}
Within this convention, negative values of $q$ correspond to prolate
configurations, whereas positive values describe oblate deformations. Since
the physical monopole mass differs from the metric parameter whenever
$q\neq0$, both normalizations are retained throughout the paper. Quantities
expressed in units of $M$ describe the geometry in its natural coordinates,
while normalization by $\mathcal{M}$ provides the corresponding scale at
fixed physical mass.

The $q$-metric has been widely employed as a simple framework for studying
the influence of static quadrupolar deformations on relativistic observables,
including multipole moments, geodesic motion, circular-orbit stability,
shadow formation, and accretion-disc properties
\citep{Zipoy1966,Voorhees1970,Quevedo1989,Quevedo2011,Boshkayev2016,Abishev2015,Boshkayev2021,ArrietaVillamizar2021,Shaikh2022}.
Here we focus on a more specific problem: the geometrical structure of
equatorial disc crossings generated by backward-traced null geodesics. The
purpose of this section is therefore to introduce the characteristic
strong-field scales that govern the transfer-function branches analysed in the
following sections.

\subsection{Equatorial optical radius and photon critical scale}
\label{subsec:optical_radius}

The higher-order transfer-function branches are produced by photons that
propagate close to unstable circular null orbits. It is therefore useful to
identify the corresponding analytic photon scale before constructing the
numerical transfer functions.

On the equatorial plane, $\theta=\pi/2$, the spacetime admits two conserved quantities, the photon energy
$E$ and axial angular momentum $L$, with impact parameter
$b=L/E$. The relevant optical radius is
\begin{equation}
h_q^2(r)
\equiv
-\frac{g_{\phi\phi}}{g_{tt}}
=
r^2
\left(1-\frac{2M}{r}\right)^{-(1+2q)} ,
\label{eq:optical_radius_q}
\end{equation}
and the circular null orbit follows from
\begin{equation}
\frac{d h_q^2}{dr}=0.
\label{eq:photon_condition_h}
\end{equation}

This condition gives
\begin{equation}
r_{\rm ph}=M(3+2q),
\label{eq:rph}
\end{equation}
which reduces to the Schwarzschild photon-sphere radius $r_{\rm ph}=3M$ when $q=0$. In the $q$-metric this orbit also marks the innermost limit of the equatorial timelike circular-orbit family and therefore defines the characteristic strong-field scale for the transfer-function analysis.

The corresponding critical impact parameter is
\begin{equation}
b_{\rm cr}(q)
=
h_q(r_{\rm ph})
=
M(3+2q)
\left(
\frac{3+2q}{1+2q}
\right)^{(1+2q)/2},
\label{eq:bcrit_q}
\end{equation}
or, equivalently,
\begin{equation}
\frac{b_{\rm cr}(q)}{M}
=
(3+2q)^{(3+2q)/2}
(1+2q)^{-(1+2q)/2}.
\label{eq:bcrit_q_equiv}
\end{equation}

For $q=0$,
\begin{equation}
b_{\rm cr}(0)=3\sqrt3\,M.
\label{eq:bcrit_schw}
\end{equation}

The photon orbit remains outside the singular surface only if
\begin{equation}
r_{\rm ph}>2M
\quad\Longleftrightarrow\quad
q>-\frac12,
\label{eq:q_domain_photon_outside}
\end{equation}
which simultaneously guarantees that the critical impact parameter is real and
positive. Throughout this work we therefore consider the moderate deformation
range
\begin{equation}
q=-0.3,\,-0.2,\,0,\,0.2,\,0.3,
\label{eq:q_values}
\end{equation}
well inside the allowed domain.

A useful analytical reference follows from the monotonic behaviour of the
critical impact parameter. Taking the logarithm of
Eq.~\eqref{eq:bcrit_q_equiv} gives
\begin{equation}
\ln\left(\frac{b_{\rm cr}}{M}\right)
=
\frac{3+2q}{2}\ln(3+2q)
-
\frac{1+2q}{2}\ln(1+2q),
\label{eq:log_bcrit}
\end{equation}
from which
\begin{equation}
\frac{d}{dq}\ln b_{\rm cr}(q)
=
\ln
\left(
\frac{3+2q}{1+2q}
\right)
\label{eq:bcrit_monotonicity}
\end{equation}
immediately follows. Since the logarithm is positive for
$q>-1/2$,

\begin{equation}
\frac{db_{\rm cr}}{dq}>0.
\label{eq:bcrit_increasing}
\end{equation}

The critical photon scale therefore increases monotonically with the
quadrupolar deformation. This behaviour provides the principal analytical
reference against which the numerical displacement of the higher-order
transfer-function branches will be compared.

\subsection{Circular-orbit hierarchy and disc-edge interpretation}
\label{subsec:circular_hierarchy}

The transfer functions are determined by photon trajectories, but the emitting
matter follows timelike circular orbits. We therefore briefly discuss the
relevant circular-orbit radii before specifying the disc model used later.

For a massive particle moving on the equatorial plane, the effective potential
can be written as
\begin{equation}
V_{\rm eff}^2(r,q)
=
\left(1-\frac{2M}{r}\right)^{q+1}
\left[
\frac{\ell^2}{r^2}
\left(1-\frac{2M}{r}\right)^q
+1
\right],
\label{eq:veff_timelike}
\end{equation}
where $\ell$ is the specific angular momentum. The circular-orbit condition
then gives
\begin{equation}
\ell^2
=
\frac{
M(1+q)r^2
\left(1-\frac{2M}{r}\right)^{-q}
}
{
r-M(3+2q)
}.
\label{eq:ell_circular}
\end{equation}

For the exterior region $r>2M$, the denominator is positive only when

\begin{equation}
r>M(3+2q)=r_{\rm ph}.
\label{eq:circular_allowed_domain}
\end{equation}

Thus, the equatorial photon orbit sets the lower radial limit of the timelike
circular-orbit family.

The marginal-stability condition has two formal solutions,
\begin{equation}
r_{\rm LSCO}^{\pm}
=
M
\left(
4+3q
\pm
\sqrt{5q^2+10q+4}
\right).
\label{eq:rlsco_pm}
\end{equation}

Only the outer solution is relevant for the values of $q$ considered in this
work:
\begin{equation}
r_{\rm LSCO}^{+}
=
M
\left(
4+3q
+
\sqrt{5q^2+10q+4}
\right).
\label{eq:rlsco_plus}
\end{equation}

At $q=0$, this reduces to the Schwarzschild result
$r_{\rm LSCO}^{+}=6M$. The distance between the outer marginally stable orbit
and the photon orbit is
\begin{equation}
r_{\rm LSCO}^{+}-r_{\rm ph}
=
M
\left[
1+q+\sqrt{5q^2+10q+4}
\right].
\label{eq:rlsco_minus_rph}
\end{equation}

This quantity is positive over the full parameter range used below. Hence,
\begin{equation}
2M<r_{\rm ph}<r_{\rm LSCO}^{+}
\label{eq:radius_hierarchy}
\end{equation}
for all models included in the calculation.

The inner boundary adopted in the radiative section is
$r_{\rm in}/\mathcal M=6.2$, where $\mathcal M=(1+q)M$ is the physical
monopole mass. The same value is used for every $q$, so the emitting region
does not acquire an additional deformation dependence through a changing
inner edge. This radius also remains outside the outer marginally stable orbit.
The largest value of $r_{\rm LSCO}^{+}/\mathcal M$ in the parameter set occurs
at $q=-0.3$ and is approximately $6.149$. It is smaller for the other values
of $q$. The adopted inner boundary therefore lies in the stable
circular-orbit region throughout the radiative calculation.

We limit the analysis to the moderate deformation interval specified in
Eq.~\eqref{eq:q_values}. For more negative values of $q$, the circular-orbit
structure near the singular surface may contain additional regions of stable
motion. These cases are outside the scope of the present study.

\subsection{Mass normalization and observable scale}
\label{subsec:mass_normalization}

The distinction between the metric parameter $M$ and the physical monopole
mass $\mathcal M$ plays an important role in the interpretation of the
numerical results. Because 
\begin{equation}
\mathcal M=(1+q)M,
\end{equation}
the same impact parameter can be expressed either in the natural coordinates
of the metric or in units of the physical mass relevant for observational
comparisons. 

Expanding ~Eq.\eqref{eq:bcrit_q_equiv} about the Schwarzschild limit gives
\begin{equation}
    \frac{b_{\rm cr}}{\mathcal M}
    =
    3\sqrt{3}
    \left[
        1+(\ln 3-1)q
        +
        \left(
            \frac{1}{2}(\ln 3-1)^2-\frac{1}{6}
        \right)q^2
        +O(q^3)
    \right].
    \label{eq:appbcrit_q_equiv}
\end{equation}
Since $\ln 3-1\simeq0.09861>0$, the critical scale initially
increases with positive $q$ at fixed $\mathcal M$. The smallness
of this coefficient also demonstrates analytically that the residual
deformation dependence is substantially weaker than the variation
expressed in metric-parameter units.

For an arbitrary impact parameter $b(q)$, we define 
\begin{equation}
\widehat b(q)
=
\frac{b(q)}{\mathcal M}
=
\frac{b(q)}{(1+q)M},
\label{eq:physical_mass_normalized_b}
\end{equation}
and, in particular,
\begin{equation}
\widehat b_{\rm cr}(q)
=
\frac{b_{\rm cr}(q)}{(1+q)M},
\qquad
\widehat b_{3,c}(q)
=
\frac{b_{3,c}(q)}{(1+q)M}.
\label{eq:normalized_bcrit_b3c}
\end{equation}

Throughout the paper, results are presented in both normalizations. The
coordinate normalization, $b/M$, is convenient for analysing the intrinsic
properties of the metric, whereas $b/\mathcal M$ removes the change of the
physical monopole mass and therefore provides a more meaningful comparison
between spacetimes with different quadrupolar deformations. As will be shown
later, this distinction substantially reduces the apparent displacement of the
higher-order transfer-function branch.

\subsection{Analytic reference table}
\label{subsec:analytic_reference_table}

Table~\ref{tab:analytic_scales} summarizes the characteristic strong-field
scales for the deformation parameters used throughout the ray-tracing
calculations. The monotonic increase of the critical impact parameter
$b_{\rm cr}/M$ follows directly from
Eq.~\eqref{eq:bcrit_increasing}, whereas the corresponding quantity
$b_{\rm cr}/\mathcal M$ varies much more slowly because the dominant
monopole rescaling has been removed.

\begin{table}
\centering
\caption{Analytic strong-field scales of the q-metric for the deformation parameters used in the ray-tracing calculation. The last column gives the critical impact parameter normalized by the physical monopole mass $\mathcal{M}=(1+q)M$.}
\label{tab:analytic_scales}
\begin{tabular}{cccccc}
\hline
\(q\) &
\(r_{\rm ph}/M\) &
\(r_{\rm LSCO}^{+}/M\) &
\((r_{\rm LSCO}^{+}-r_{\rm ph})/M\) &
\(b_{\rm cr}/M\) &
\(b_{\rm cr}/\mathcal{M}\) \\
\hline
\(-0.3\) & 2.4000 & 4.3042 & 1.9042 & 3.4343 & 4.9062 \\
\(-0.2\) & 2.6000 & 4.8832 & 2.2832 & 4.0366 & 5.0458 \\
\(0\)    & 3.0000 & 6.0000 & 3.0000 & 5.1962 & 5.1962 \\
\(0.2\)  & 3.4000 & 7.0900 & 3.6900 & 6.3274 & 5.2728 \\
\(0.3\)  & 3.6000 & 7.6295 & 4.0295 & 6.8872 & 5.2979 \\
\hline
\end{tabular}
\end{table}

The analytic critical impact parameter and the numerical third-branch centre
characterize the same strong-field region but are not identical quantities.
The former is defined by the unstable circular null orbit, whereas the latter
is obtained as the midpoint of a finite interval of rays producing a third
equatorial disc crossing. Their absolute values therefore need not coincide.
Instead, the comparison focuses on their common monotonic dependence on the
quadrupolar deformation, providing an analytical reference for the numerical
transfer-function diagnostics presented in the following sections.

\section{Ray tracing and disc-crossing algorithm}
\label{sec:raytracing}

We obtain the disc-to-screen mapping by tracing photon trajectories backwards
from a distant observer through the fixed $q$-metric spacetime. At this stage, each ray is characterized only by its geodesic trajectory and by the
ordered radii at which it intersects the equatorial plane. Emissivity and
frequency-shift factors are introduced separately in Sec.~\ref{sec:emission}.

Photon motion is governed by the Hamiltonian
\begin{equation}
H=\frac{1}{2}g^{\mu\nu}p_\mu p_\nu=0,
\label{eq:Hamiltonian}
\end{equation}
with Hamilton equations
\begin{equation}
\frac{dx^\mu}{d\lambda}
=
\frac{\partial H}{\partial p_\mu},
\qquad
\frac{dp_\mu}{d\lambda}
=
-\frac{\partial H}{\partial x^\mu},
\label{eq:Hamilton_equations}
\end{equation}
where $\lambda$ is an affine parameter. Since the spacetime is stationary
and axisymmetric, $p_t$ and $p_\phi$ are conserved. For the $+---$
signature, we define
\begin{equation}
E=p_t,
\qquad
L_z=-p_\phi .
\label{eq:conserved_quantities}
\end{equation}

For the nearly face-on transfer-function calculation, the observer is placed
at
\begin{equation}
r_{\rm obs}=1000M,
\qquad
\theta_{\rm obs}=10^{-4}.
\label{eq:observer_faceon}
\end{equation}

The observer is static in the coordinates of Eq.~\eqref{eq:q_metric}. At the observer position, we introduce the orthonormal tetrad
\begin{align} \nonumber
& e_{(\hat t)}^{\ \mu}
=
\frac{\delta^\mu_t}{\sqrt{g_{tt}}},
\quad
e_{(\hat r)}^{\ \mu}
=
\frac{\delta^\mu_r}{\sqrt{-g_{rr}}},
\quad
e_{(\hat\theta)}^{\ \mu}
=
\frac{\delta^\mu_\theta}{\sqrt{-g_{\theta\theta}}},
\\
& e_{(\hat\phi)}^{\ \mu}
=
\frac{\delta^\mu_\phi}{\sqrt{-g_{\phi\phi}}},
\label{eq:static_tetrad}
\end{align}
where all metric components are evaluated at
$(r_{\rm obs},\theta_{\rm obs})$.

The observer screen is the local plane orthogonal to the radial tetrad
direction. We use Cartesian screen coordinates $(X,Y)$, with $X$ directed
along $+e_{(\hat\phi)}$ and positive $Y$ along $-e_{(\hat\theta)}$. A photon launched backwards from the screen point $(X,Y)$ is assigned the local momentum
\begin{align} \nonumber
& p^{(\hat t)}=1,
\quad
p^{(\hat r)}=-\frac{1}{\mathcal N},
\quad
p^{(\hat\theta)}
=
-\frac{Y/r_{\rm obs}}{\mathcal N},
\\
& p^{(\hat\phi)}
=
\frac{X/r_{\rm obs}}{\mathcal N},
\label{eq:general_screen_direction}
\end{align}
where
\begin{equation}
\mathcal N
=
\sqrt{1+\frac{X^2+Y^2}{r_{\rm obs}^2}} .
\label{eq:general_screen_normalization}
\end{equation}
This choice satisfies the local null condition. The coordinate-basis
momentum and the covariant Hamiltonian variables then follow from
\begin{equation}
p^\mu=e_{(\hat a)}^{\ \mu}p^{(\hat a)},
\qquad
p_\mu=g_{\mu\nu}p^\nu .
\label{eq:tetrad_to_coordinate_momentum}
\end{equation}

For the one-dimensional nearly face-on calculation, axial symmetry allows us
to set $X=0$. The rays then lie in a meridional plane and satisfy $p^{(\hat\phi)}=0$. Introducing the local launch angle $\alpha$, their initial momentum can be written as
\begin{equation}
p^{(\hat t)}=1,
\quad
p^{(\hat r)}=-\cos\alpha,
\quad
p^{(\hat\theta)}=\sin\alpha,
\quad
p^{(\hat\phi)}=0.
\label{eq:tetrad_momentum}
\end{equation}
We label this ray family by the finite-distance impact parameter
\begin{equation}
b=r_{\rm obs}\sin\alpha .
\label{eq:impact_parameter}
\end{equation}
For the screen orientation defined above,
\begin{equation}
b
=
-\frac{Y}
{\sqrt{1+Y^2/r_{\rm obs}^2}}
\qquad (X=0).
\label{eq:b_screen_relation}
\end{equation}
The minus sign is a consequence of the convention
$Y\parallel-e_{(\hat\theta)}$. In the small-angle limit,
$|Y|\ll r_{\rm obs}$, one has $b\simeq-Y$. At large observer radius,
$b$ approaches the usual asymptotic screen impact parameter. It should not
be confused with $L_z/E$, because the meridional rays considered here have
$p_\phi=0$.

The disc is represented by an infinitesimally thin equatorial surface. A
crossing is recorded when a backward-integrated trajectory satisfies
\begin{equation}
\cos\theta(\lambda)=0,
\label{eq:disc_crossing_condition}
\end{equation}
provided that the crossing radius lies outside the numerical cutoff near
$r=2M$. This construction is the geometrical basis of relativistic
thin-disc transfer functions
\citep{Cunningham1975,Luminet1979}.

When a photon intersects the equatorial plane several times, the crossing
radii are ordered by increasing affine parameter:
\begin{equation}
\{r_1,r_2,r_3,\ldots\}
=
\{r_1(b;q),r_2(b;q),r_3(b;q),\ldots\}.
\label{eq:crossing_sequence}
\end{equation}
We retain the first three intersections. The first crossing describes the
direct branch, the second gives the lensed-ring branch, and the third
corresponds to the first photon-ring-like branch generated by near-critical
propagation. These branches are defined by crossing order; their radiative
weight in an image is considered later.

The same Hamiltonian system and screen construction are used for the
two-dimensional calculations in Sec.~\ref{sec:diagnostics}. In that case,
both $X$ and $Y$ are varied, and each screen point is assigned the radius
of its first equatorial crossing,
\begin{equation}
r_{\rm hit}=r_{\rm hit}(X,Y;q).
\label{eq:rhit_definition}
\end{equation}
The function $r_{\rm hit}$ is a first-intersection map. It can be used
directly to extract iso-radial contours or combined with an emission model
and a redshift factor to construct a direct image. The map itself contains
no intensity information.

The stopping conditions, crossing refinement, screen resolution, and
convergence tests are described in ~\ref{app:numerics}. The ordered
crossing sequence in Eq.~\eqref{eq:crossing_sequence} provides the transfer
functions defined below.

\section{Transfer functions}
\label{sec:transfer}

For each nearly face-on screen impact parameter $b$, the ray-tracing
procedure of Sec.~\ref{sec:raytracing} returns an ordered set of equatorial
crossing radii,
\begin{equation}
{r_1,r_2,r_3,\ldots}
= {r_1(b;q),r_2(b;q),r_3(b;q),\ldots}.
\label{eq:transfer_functions}
\end{equation}
The functions $r_m(b;q)$ are labelled by the order in which the
backward-traced photon intersects the disc. The first branch
$r_1(b;q)$ maps the direct disc image. The second branch $r_2(b;q)$
describes the lensed-ring contribution, while $r_3(b;q)$ is the first
higher-order branch produced by trajectories that pass close to the
near-critical region. We refer to the latter as the photon-ring-like branch.

{In the present work, the term transfer function refers specifically to the geometrical mapping between the observer-screen impact parameter and the radius of the $m$th equatorial disc crossing. Radiative weights, including the emissivity and frequency-shift factors, are introduced separately and are not part of the definition of $r_m(b;q)$.}

The transfer functions specify where a ray reaches the disc, but not how
bright that crossing appears to the observer. An emissivity profile, the
emitting radial range, and the emitter-to-observer frequency shift are needed
to assign an observed intensity. These ingredients weight the branches
without changing their position in the $(b,r_m)$ plane. In this sense, the
transfer functions provide the geometrical input to the radiative
calculation introduced in Sec.~\ref{sec:emission}
\citep{Cunningham1975,Luminet1979,Gralla2019}.

Before examining the deformed cases, we validate the numerical construction
in the Schwarzschild limit. For $q=0$, the analytic critical impact
parameter is
\begin{equation}
b_{\rm cr}(0)=3\sqrt{3} M\simeq 5.196 M,
\label{eq:transfer_schw_bcrit}
\end{equation}
as obtained in Eq.~\eqref{eq:bcrit_schw}. The resolved numerical interval of
the third branch is
\begin{equation}
5.1832\lesssim \frac{b}{M}\lesssim 5.2222 .
\label{eq:schw_validation}
\end{equation}
It contains the critical value $3\sqrt{3}$, confirming that the
third-crossing branch samples the expected near-critical Schwarzschild
region. This validation concerns only the geodesic crossing structure and
does not depend on the subsequent radiative prescription.

For $q\neq0$, the analytic critical impact parameter
$b_{\rm cr}(q)$ is given by Eq.~\eqref{eq:bcrit_q}. It is determined by
the unstable equatorial circular null orbit. The numerical branch
$r_3(b;q)$, by contrast, consists of a finite interval of rays that produce
a third equatorial crossing. The two constructions probe the same
near-critical region, but they need not give identical impact parameters.

To compare them, we introduce the midpoint of the resolved third-branch
interval,
\begin{equation}
b_{3,c}(q)
=
\frac{1}{2}
\left[
b_{3,\min}(q)+b_{3,\max}(q)
\right],
\label{eq:b3c_definition}
\end{equation}
where $b_{3,\min}$ and $b_{3,\max}$ are its numerical endpoints. The
quantity $b_{3,c}$ is an operational branch diagnostic. It is not an exact
critical parameter and is not used as a substitute for $b_{\rm cr}$.

Table~\ref{tab:bcr_validation} lists the two scales in units of both the
metric parameter $M$ and the physical monopole mass
$\mathcal{M}=M(1+q)$. The Schwarzschild values differ only slightly. Away
from $q=0$, the finite third-crossing interval is displaced relative to
the analytic limiting orbit, and the sign and magnitude of this offset vary
with the deformation. Section~\ref{sec:tf_results} examines this behaviour
together with the full branch structure.

\begin{table}
\centering
\caption{Analytic critical impact parameter and numerical third-branch
centre. Here $\mathcal{M}=(1+q)M$ is the physical monopole mass.}
\label{tab:bcr_validation}
\begin{tabular}{ccccc}
\hline
\(q\) & \(b_{\rm cr}/M\) & \(b_{3,c}/M\) &
\(b_{\rm cr}/\mathcal{M}\) & \(b_{3,c}/\mathcal{M}\) \\
\hline
\(-0.3\) & 3.4343 & 3.7163 & 4.9062 & 5.3090 \\
\(-0.2\) & 4.0366 & 4.2094 & 5.0458 & 5.2617 \\
0        & 5.1962 & 5.2027 & 5.1962 & 5.2027\\
0.2      & 6.3274 & 6.2016 & 5.2728 & 5.1680 \\
0.3      & 6.8873 & 6.7025 & 5.2979 & 5.1558 \\
\hline
\end{tabular}
\end{table}

\section{Face-on transfer-function results}
\label{sec:tf_results}

We now apply the construction of Secs.~\ref{sec:raytracing} and
\ref{sec:transfer} to the deformation parameters listed in
Eq.~\eqref{eq:q_values}. For each $q$, the ordered crossing radii are
extracted as functions of the nearly face-on screen impact parameter $b$.
The first three resolved branches are denoted by $r_1$, $r_2$, and
$r_3$. Their classification refers to crossing order: $r_1$ is the
direct branch, $r_2$ the lensed-ring branch, and $r_3$ the first
photon-ring-like branch.

Table~\ref{tab:ranges} gives the resolved impact-parameter intervals. The
direct branch extends over a broad range because it includes rays that reach
the disc after comparatively weak deflection. The second and third branches
are confined to the strong-lensing region. In particular, $r_3$ occupies
a narrow interval close to the analytic photon critical scale derived in
Sec.~\ref{subsec:optical_radius}.

\begin{table}
\centering
\caption{Resolved impact-parameter intervals of the first three
disc-crossing branches. The analytic photon-orbit radius $r_{\rm ph}$ and
the outer marginally stable circular-orbit radius $r_{\rm LSCO}^{+}$ are
included as reference strong-field scales. The quoted branch endpoints are
defined by the numerical sampling and crossing-selection procedure described
in Sec.~\ref{sec:raytracing}. For the direct branch, the upper endpoint corresponds to the
adopted scan limit and should not be interpreted as a physical
termination of the branch.}
\label{tab:ranges}
\begin{tabular}{cccccc}
\hline
$q$ &
$r_{\rm ph}/M$ &
$r_{\rm LSCO}^{+}/M$ &
$r_1: b/M$ &
$r_2: b/M$ &
$r_3: b/M$ \\
\hline
$-0.3$ & $2.4000$ & $4.3042$ & $2.04$--$12.0$ & $3.6093$--$4.3393$ & $3.7063$--$3.7263$ \\
$-0.2$ & $2.6000$ & $4.8832$ & $2.35$--$12.0$ & $4.0766$--$4.9416$ & $4.1961$--$4.2226$ \\
$0$    & $3.0000$ & $6.0000$ & $2.85$--$12.0$ & $5.0111$--$6.1461$ & $5.1832$--$5.2222$ \\
$0.2$  & $3.4000$ & $7.0900$ & $3.40$--$12.0$ & $5.9574$--$7.3574$ & $6.1754$--$6.2279$ \\
$0.3$  & $3.6000$ & $7.6295$ & $3.70$--$12.0$ & $6.4273$--$7.9623$ & $6.6733$--$6.7318$ \\
\hline
\end{tabular}
\end{table}

\begin{figure}
\centering
\includegraphics[width=0.55\columnwidth]
{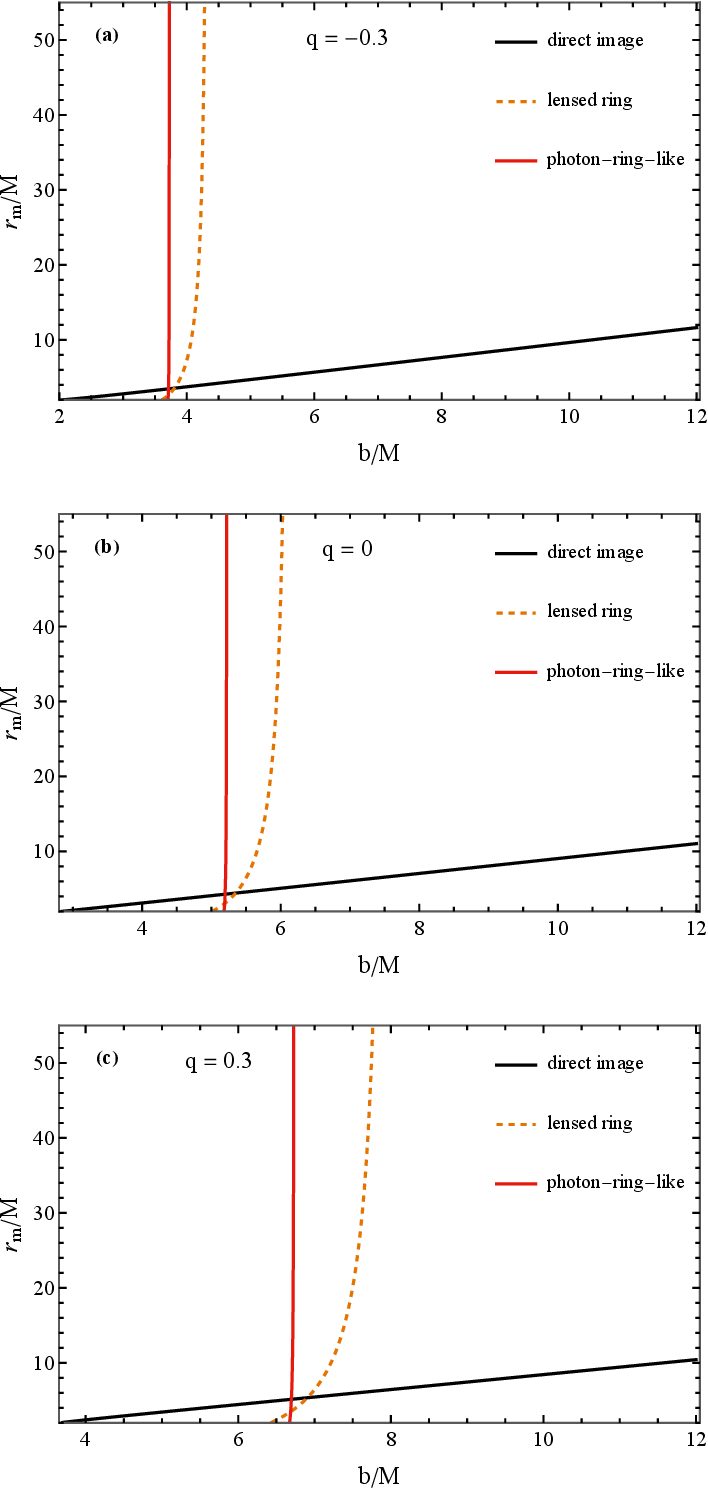}
\caption{First three disc-crossing transfer curves for
$q=-0.3$, $q=0$, and $q=0.3$. The black, orange, and red curves denote
the first-, second-, and third-crossing branches $r_1$, $r_2$, and
$r_3$, respectively. The higher-order branches are concentrated near the
strong-lensing region and shift towards larger $b/M$ as $q$ increases.}
\label{fig:three_branches}
\end{figure}

Representative transfer curves are shown in
Fig.~\ref{fig:three_branches} for $q=-0.3$, $q=0$, and $q=0.3$.
The direct branch changes smoothly with $b$, whereas the higher-order
branches rise sharply within more restricted intervals. In the coordinate
normalization set by $M$, the second and third branches both move towards
larger impact parameters as $q$ increases.

\FloatBarrier
To quantify the resolved third branch, we use the centre $b_{3,c}$ defined
in Eq.~\eqref{eq:b3c_definition}. Its numerical width is
\begin{equation}
\Delta b_3(q)
=
b_{3,\max}(q)-b_{3,\min}(q).
\label{eq:delta_b3_definition}
\end{equation}
We also define its displacement from the Schwarzschild value and the
corresponding centre ratio,
\begin{equation}
\delta b_{3,c}(q)
=
b_{3,c}(q)-b_{3,c}(0),
\qquad
\eta_3(q)
=
\frac{b_{3,c}(q)}{b_{3,c}(0)}.
\label{eq:r3_relative_diagnostics}
\end{equation}
These quantities refer to a finite numerically resolved interval. They are
not invariant photon-ring observables, and their endpoint values depend on
the common sampling and branch-selection procedure described in
Sec.~\ref{sec:raytracing}. Their purpose is to compare the branch location
and width in the same set of numerical calculations.

\begin{figure}
\centering
\includegraphics[width=0.52\columnwidth]
{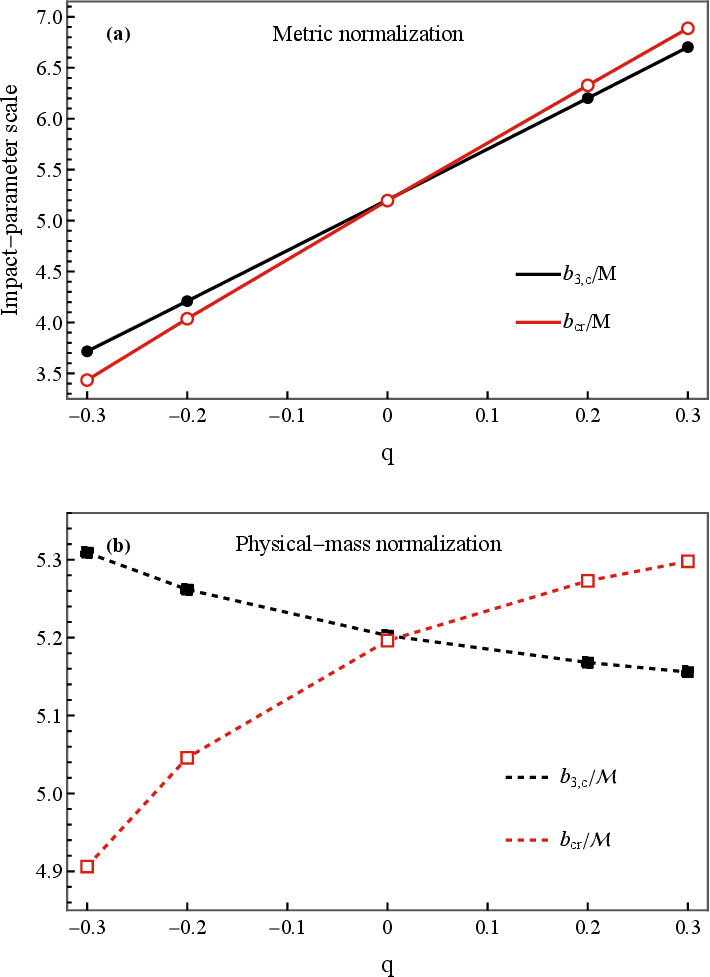}
\caption{Third-branch centre $b_{3,c}$ (black color) compared with the analytic photon
critical scale $b_{\rm cr}$ (red color). The upper panel uses the metric normalization,
whereas the lower panel uses the physical monopole mass
$\mathcal{M}=M(1+q)$. The strong displacement visible in units of $M$ is
substantially reduced when the physical mass is held fixed, leaving a smaller
residual dependence on the quadrupolar deformation.}
\label{fig:b3_bcrit_normalization}
\end{figure}

Table~\ref{tab:r3diagnostic} lists the resulting diagnostics, including
$\Delta b_3$ and the normalized physical-mass offset
$(b_{3,c}-b_{\rm cr})/\mathcal{M}$.
Figure~\ref{fig:b3_bcrit_normalization} compares $b_{3,c}$ with the
analytic critical scale in the two mass normalizations used throughout the
paper.

\begin{table*}
\centering
\caption{Diagnostics of the resolved third-crossing branch. The centre
$b_{3,c}$ is the midpoint of the numerically resolved interval
$[b_{3,\min},b_{3,\max}]$, and
$\Delta b_3=b_{3,\max}-b_{3,\min}$ is its resolved width.
The final three columns compare the analytic critical scale, the numerical
branch centre, and their difference after normalization by the physical
monopole mass $\mathcal{M}=(1+q)M$.}
\label{tab:r3diagnostic}
\begin{tabular}{ccccccccccc}
\hline
$q$ &
$r_{\rm ph}/M$ &
$b_{3,\min}/M$ &
$b_{3,\max}/M$ &
$\Delta b_3/M$ &
$b_{3,c}/M$ &
$\delta b_{3,c}/M$ &
$\eta_3$ &
$b_{\rm cr}/\mathcal{M}$ &
$b_{3,c}/\mathcal{M}$ &
$(b_{3,c}-b_{\rm cr})/\mathcal{M}$ \\
\hline
$-0.3$ &
$2.4000$ &
$3.7063$ &
$3.7263$ &
$0.0200$ &
$3.7163$ &
$-1.4863$ &
$0.7143$ &
$4.9062$ &
$5.3090$ &
$0.4029$ \\

$-0.2$ &
$2.6000$ &
$4.1961$ &
$4.2226$ &
$0.0265$ &
$4.2094$ &
$-0.9933$ &
$0.8091$ &
$5.0458$ &
$5.2617$ &
$0.2159$ \\

$0$ &
$3.0000$ &
$5.1832$ &
$5.2222$ &
$0.0390$ &
$5.2027$ &
$0.0000$ &
$1.0000$ &
$5.1962$ &
$5.2027$ &
$0.0065$ \\

$0.2$ &
$3.4000$ &
$6.1754$ &
$6.2279$ &
$0.0525$ &
$6.2016$ &
$0.9990$ &
$1.1920$ &
$5.2728$ &
$5.1680$ &
$-0.1048$ \\

$0.3$ &
$3.6000$ &
$6.6733$ &
$6.7318$ &
$0.0585$ &
$6.7025$ &
$1.4999$ &
$1.2883$ &
$5.2979$ &
$5.1558$ &
$-0.1421$ \\
\hline
\end{tabular}

\end{table*}

In units of the metric parameter, the third-branch centre changes from
$b_{3,c}/M\simeq3.716$ at $q=-0.3$ to
$b_{3,c}/M\simeq6.702$ at $q=0.3$. These values are approximately
$28.6\%$ below and $28.8\%$ above the Schwarzschild result,
respectively. The monotonic ordering is a property of the null-geodesic
crossing map. It is therefore already established before the same map is
weighted by an emissivity profile and a frequency-shift factor in the
radiative calculation.

The resolved interval also broadens with increasing $q$:
$\Delta b_3/M$ rises from approximately $0.0200$ at $q=-0.3$ to
$0.0585$ at $q=0.3$. This is a reproducible trend for the sampling
prescription used here, but $\Delta b_3$ should not be interpreted as a
physical photon-ring width. Its value is tied to the numerical identification
of the finite third-crossing interval and is consequently more sensitive to
resolution than its centre.

The apparent displacement is much smaller when the physical monopole mass is
held fixed. In this normalization,
\[
\frac{b_{3,c}}{\mathcal M}
\simeq5.309,\ 5.262,\ 5.203,\ 5.168,\ 5.156
\]
for $q=-0.3,-0.2,0,0.2,0.3$, respectively. Thus,
$b_{3,c}/\mathcal M$ decreases mildly over the sampled interval, even
though $b_{3,c}/M$ increases strongly. The contrast between the two
normalizations shows that most of the large shift in coordinate units is
associated with the relation $\mathcal M=M(1+q)$, rather than with a
comparably large residual change at fixed physical mass.

The finite third branch does not remain centred exactly on the analytic
critical scale. Its offset is
\[
\frac{b_{3,c}-b_{\rm cr}}{\mathcal{M}}
\simeq0.403,\ 0.216,\ 0.0065,\ -0.105,\ -0.142
\]
for $q=-0.3,-0.2,0,0.2,0.3$, respectively. The centre lies above
$b_{\rm cr}$ for negative $q$, nearly coincides with it in the
Schwarzschild case, and lies below it for positive $q$. This sign change
does not indicate a failure of the ray tracing. The two quantities have
different definitions: $b_{\rm cr}$ belongs to the limiting unstable
circular null orbit, whereas $b_{3,c}$ is the midpoint of a finite set of
rays that complete a third disc crossing.

The comparison therefore leads to two separate conclusions. First, the
resolved higher-order branch follows the strong displacement of the analytic
critical scale when lengths are measured in units of $M$. Second, after
normalization by $\mathcal M$, the remaining variation is smaller and is
not simply a monotonic enlargement of the lensing scale. The latter is the
more relevant result for comparisons made at fixed physical monopole mass.
Sections\ref{sec:diagnostics} and ~\ref{sec:emission} use this
fixed-$\mathcal M$ viewpoint to examine how the residual deformation enters
an inclined screen map and its minimally weighted direct image.


\section{Inclined first-intersection maps and contour diagnostics}
\label{sec:diagnostics}

The transfer functions presented in Sec.~\ref{sec:tf_results} describe the
successive equatorial intersections of photon trajectories for a nearly
face-on observer. They provide a one-dimensional view of the image structure,
but do not show how the disc is mapped across an inclined observer screen. We
therefore complement that analysis with two-dimensional maps constructed at a
finite viewing angle. In this calculation, only the first equatorial
intersection of each ray is retained.

The ray tracing uses the Hamiltonian equations, static-observer tetrad, and
screen-to-momentum prescription introduced in
Sec.~\ref{sec:raytracing}. For the present calculation, the rays are launched
from a two-dimensional screen and the observer is placed at

\begin{equation}
\frac{r_{\rm obs}}{\mathcal M}=1000,
\qquad
\theta_{\rm obs}=80^\circ,
\label{eq:inclined_observer_position}
\end{equation}
where $\theta_{\rm obs}$ is measured from the symmetry axis. Thus,
$\theta_{\rm obs}=0^\circ$ corresponds to a face-on observer, whereas
$\theta_{\rm obs}=90^\circ$ gives an edge-on view.

The screen coordinates $(X,Y)$ are defined as in Sec.~\ref{sec:raytracing}: the horizontal direction is aligned with the local azimuthal basis vector and the vertical direction with the local polar basis vector. All comparisons in this section are made at fixed physical monopole mass,
\begin{equation}
\mathcal M=(1+q)M.
\end{equation}

The metric parameter is therefore
\begin{equation}
M=\frac{\mathcal M}{1+q},
\end{equation}
while the observer position, screen coordinates, and reference disc radii are
kept fixed in units of $\mathcal M$. This normalization removes the overall
monopole-mass rescaling from the comparison between different values of $q$.

For each accepted ray, the integration records the radius of its first
intersection with the equatorial plane,
\begin{equation}
r_{\rm hit}=r_{\rm hit}(X,Y;q).
\label{eq:rhit_map}
\end{equation}

This defines a two-dimensional first-intersection map on the observer screen.
A contour satisfying
\begin{equation}
r_{\rm hit}(X,Y;q)=r_0
\label{eq:isoradial_contours}
\end{equation}
is the apparent image of a circular ring of radius $r_0$ in the equatorial
plane. At this stage the map is purely geometrical: no emissivity,
frequency-shift factor, or intensity weighting is included.

We use four reference radii,
\begin{equation}
\frac{r_0}{\mathcal M}
=
6,\;
10,\;
20,\;
30,
\label{eq:contour_radii}
\end{equation}
which sample both the inner and outer parts of the disc. The additional
contour $r_{\rm hit}/\mathcal M=6.2$ is considered separately because it
coincides with the inner emitting-disc boundary used in Sec.~\ref{sec:emission}.

For each contour $C_{q,r_0}$, the horizontal and vertical extents are defined
by 
\begin{equation}
\Delta X=X_{\max}-X_{\min},
\qquad
\Delta Y=Y_{\max}-Y_{\min}.
\label{eq:contour_extents}
\end{equation}

Their ratio,
\begin{equation}
\mathcal A_c
=
\frac{\Delta X}{\Delta Y},
\label{eq:contour_aspect_ratio}
\end{equation}
is used as a simple measure of the projected contour shape. We also define
the geometric centre from the bounding coordinates,
\begin{equation}
X_c^{\rm geom}
=
\frac{X_{\max}+X_{\min}}{2},
\qquad
Y_c^{\rm geom}
=
\frac{Y_{\max}+Y_{\min}}{2}.
\label{eq:geometric_centres}
\end{equation}

These quantities are obtained directly from the extracted contours and do
not depend on the emission model introduced later.

\begin{figure*}[t]
\centering
\includegraphics[width=0.998\textwidth]
{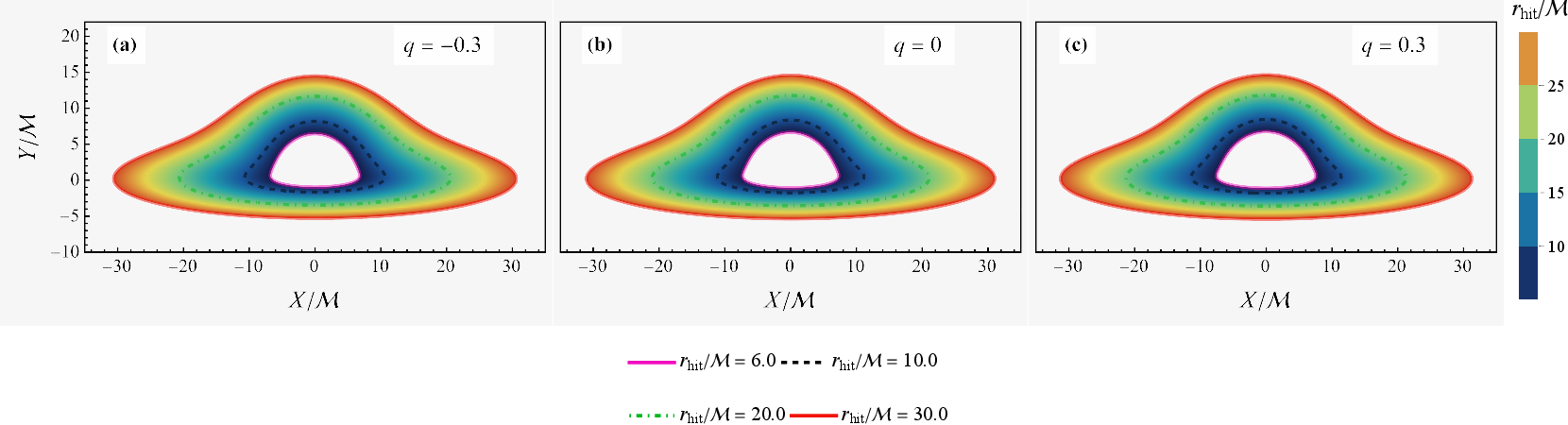}
\caption{
Inclined first-intersection radius maps for
$q=-0.3$, $q=0$, and $q=0.3$ at fixed physical monopole mass.
The overlaid curves satisfy
$r_{\rm hit}(X,Y;q)=r_0$ for
$r_0/\mathcal M=6,10,20,$ and $30$.
The colour scale gives the first equatorial-intersection radius
$r_{\rm hit}/\mathcal M$. No emissivity or frequency-shift weighting is
included. The observer is located at
$r_{\rm obs}/\mathcal M=1000$ and
$\theta_{\rm obs}=80^\circ$.
}
\label{fig:iso_contours}
\end{figure*}

\begin{table}
\centering
\caption{
Geometrical diagnostics of the inclined first-intersection contours at fixed
physical monopole mass $\mathcal M=(1+q)M$. The radii
$r_0/\mathcal M=6,10,20,$ and $30$ form the reference set shown in
Fig.~\ref{fig:iso_contours}. The additional contour
$r_0/\mathcal M=6.2$ coincides with the inner emitting-disc boundary used in
Sec.~\ref{sec:emission}. All dimensional quantities are given in units of
$\mathcal M$. No emissivity or frequency-shift weighting is included.
}
\label{tab:contour_diagnostics}
\renewcommand{\arraystretch}{1.08}
\begin{tabular}{cccccc}
\hline
$q$ &
$r_0/\mathcal M$ &
$\Delta X/\mathcal M$ &
$\Delta Y/\mathcal M$ &
$\mathcal A_c$ &
$Y_c^{\rm geom}/\mathcal M$ \\
\hline
-0.3 & 6   & 13.572 & 7.412  & 1.831 & 2.752 \\
-0.3 & 6.2 & 13.946 & 7.547  & 1.848 & 2.783 \\
-0.3 & 10  & 21.358 & 9.888  & 2.160 & 3.276 \\
-0.3 & 20  & 41.225 & 15.123 & 2.726 & 4.102 \\
-0.3 & 30  & 61.204 & 19.789 & 3.093 & 4.610 \\
\hline
0 & 6   & 14.608 & 7.703  & 1.896 & 2.803 \\
0 & 6.2 & 14.978 & 7.834  & 1.912 & 2.833 \\
0 & 10  & 22.317 & 10.126 & 2.204 & 3.309 \\
0 & 20  & 42.119 & 15.320 & 2.749 & 4.119 \\
0 & 30  & 62.071 & 19.967 & 3.109 & 4.619 \\
\hline
0.3 & 6   & 15.117 & 7.855  & 1.925 & 2.832 \\
0.3 & 6.2 & 15.495 & 7.982  & 1.941 & 2.861 \\
0.3 & 10  & 22.810 & 10.253 & 2.225 & 3.329 \\
0.3 & 20  & 42.591 & 15.425 & 2.761 & 4.129 \\
0.3 & 30  & 62.531 & 20.064 & 3.117 & 4.625 \\
\hline
\end{tabular}
\end{table}

Figure~\ref{fig:iso_contours} shows the first-intersection maps for
$q=-0.3$, $0$, and $0.3$. The contours are smooth and approximately
symmetric under $X\rightarrow-X$. No change in the topology of the displayed
maps is visible over the deformation range considered here. The main effect
of changing $q$ is a variation in the contour dimensions, accompanied by a
smaller change in their shape.

The separation between the three models is clearest in the inner part of the
disc. In particular, the contour at $r_{\rm hit}/\mathcal M=6$ changes more
noticeably with $q$ than the contours at $r_{\rm hit}/\mathcal M=10$, $20$,
and $30$. The latter become progressively closer as the radius increases,
consistent with the weakening of the quadrupolar contribution at larger
distances.

The numerical diagnostics are given in Table~\ref{tab:contour_diagnostics}. Both $\Delta X$ and $\Delta Y$ increase with $q$ at every radius listed in the table. The relative change is largest for the inner reference contour. At $r_{\rm hit}/\mathcal M=6$, the horizontal extent increases from

\begin{equation*}
\frac{\Delta X}{\mathcal M}=13.572
\end{equation*}

at $q=-0.3$ to
\begin{equation*}
\frac{\Delta X}{\mathcal M}=15.117
\end{equation*}
at $q=0.3$, corresponding to an increase of approximately $11.4\%$.
Over the same interval, $\Delta Y/\mathcal M$ increases from $7.412$ to
$7.855$, or by approximately $6.0\%$.

The changes are smaller for the outermost contour. At
$r_{\rm hit}/\mathcal M=30$, the relative increases in $\Delta X$ and
$\Delta Y$ between $q=-0.3$ and $q=0.3$ are approximately $2.2\%$ and
$1.4\%$, respectively. The deformation therefore affects the inner contours
more strongly than the outer ones.

The aspect ratio $\mathcal A_c$ also increases with $q$, although its
variation is modest. Since the horizontal and vertical extents change by
different relative amounts, the contours are not obtained from one another
by a uniform rescaling. Their projected shape changes slightly as well.

The vertical centre $Y_c^{\rm geom}$ varies only weakly across the three
models. Its positive value reflects the projected displacement of the
inclined disc rings on the screen. By contrast, reflection symmetry requires
the horizontal centre $X_c^{\rm geom}$ to vanish up to numerical error. A
direct check of this symmetry is given below.

\begin{figure}
\centering
\includegraphics[width=0.56\columnwidth]
{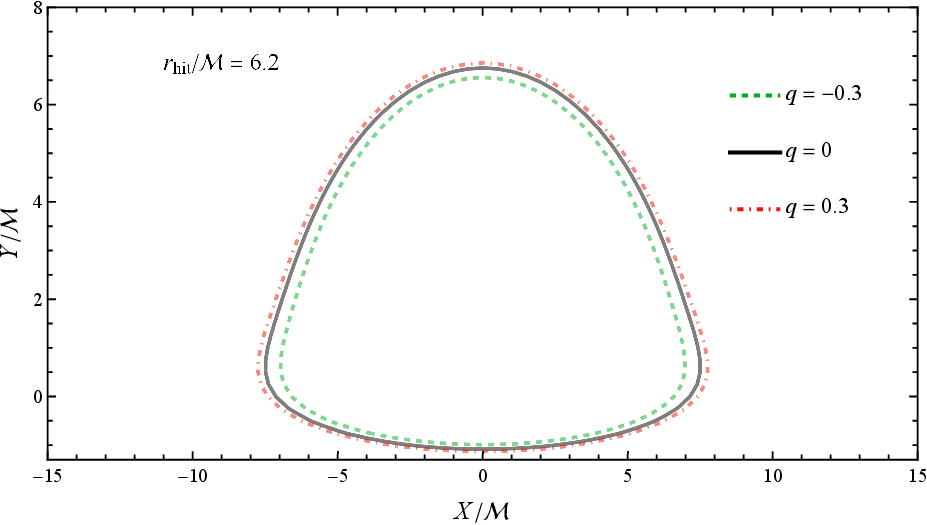}
\caption{
First-intersection contour $r_{\rm hit}/\mathcal M=6.2$ for
$q=-0.3$, $q=0$, and $q=0.3$ at fixed physical monopole mass.
This radius coincides with the inner emitting-disc boundary used in
Sec.~\ref{sec:emission} and provides a geometrical reference for the direct
image. The observer is located at
$r_{\rm obs}/\mathcal M=1000$ and
$\theta_{\rm obs}=80^\circ$.
}
\label{fig:inner_contour}
\end{figure}

The radiative calculation in the next section uses
$r_{\rm in}/\mathcal M=6.2$ as the inner disc boundary. The corresponding
first-intersection contour is therefore shown separately in
Fig.~\ref{fig:inner_contour} for $q=-0.3$, $0$, and $0.3$. The four contours
in Fig.~\ref{fig:iso_contours} are retained as the common reference set for
the broader radial comparison.

The numerical values for the $6.2$ contour are also included in
Table~\ref{tab:contour_diagnostics}. From $q=-0.3$ to $q=0.3$,
$\Delta X/\mathcal M$ increases from $13.946$ to $15.495$, while
$\Delta Y/\mathcal M$ increases from $7.547$ to $7.982$. Over the same
interval, the aspect ratio changes from $1.848$ to $1.941$. These results
follow the same trend as the nearby $r_{\rm hit}/\mathcal M=6$ contour.

We checked the left--right symmetry using full-screen maps computed without
applying the reflection $X\rightarrow-X$. For each value of $q$, the map was
constructed from a $161\times121$ grid of independently traced rays. The
largest normalized horizontal-centre residual was
\begin{equation}
\max\left(
\frac{|X_c^{\rm geom}|}{\Delta X}
\right)
=
7.3\times10^{-5}.
\label{eq:x_symmetry_residual}
\end{equation}
As a separate test, full-screen contours were reconstructed from the
$X\geq0$ half of each map by applying the same reflection. Their widths were
then compared with those obtained from the independently traced full-screen
maps for all radii listed in Table~\ref{tab:contour_diagnostics}. The largest
relative difference in $\Delta X$ was $1.464\times10^{-4}$, while no
difference was found in $\Delta Y$. Both checks are consistent with the
expected reflection symmetry at the adopted screen resolution.

No emissivity or frequency-shift weighting has been used in this section. In
Sec.~\ref{sec:emission}, an observed intensity is assigned to each accepted
first crossing using a simple emissivity law and the corresponding
frequency-shift factor. The resulting direct images can then be compared
with the contour maps obtained here.
The full-screen implementation and symmetry checks are described in
~\ref{app:inclined_maps}.

\section{Minimal radiative illustration}
\label{sec:emission}

The first-intersection maps presented in Sec.~\ref{sec:diagnostics} give the
radius at which each accepted ray first reaches the equatorial disc. They
contain no brightness information. To construct a direct image, we specify
the emission at the intersection point and the frequency shift between the
emitting matter and the observer.

We use a prescribed emissivity profile and assume that the emitting matter
follows circular equatorial geodesics. This is sufficient to examine how the
first-intersection geometry enters an intensity map, but it is not intended as
a self-consistent accretion-disc model.

The local bolometric emissivity is taken to be
\begin{equation}
I_{\rm em}(r)=
\begin{cases}
I_0\left(\dfrac{r_{\rm in}}{r}\right)^p,
& r_{\rm in}\leq r\leq r_{\rm out},\\[2mm]
0,
& \text{otherwise}.
\end{cases}
\label{eq:emissivity_profile}
\end{equation}

The normalization $I_0$ is arbitrary and cancels from the normalized
quantities used below. The power law supplies a simple radial weighting of
the disc intersections and should not be regarded as a replacement for a
self-consistent Novikov--Thorne disc \citep{PageThorne1974}.

As in Sec.~\ref{sec:diagnostics}, the comparison is performed at fixed
physical monopole mass $\mathcal M=(1+q)M$. The emitting region is the same
for all three deformations:
\begin{equation}
\frac{r_{\rm in}}{\mathcal M}=6.2,
\qquad
\frac{r_{\rm out}}{\mathcal M}=30,
\qquad
p=3.
\label{eq:emission_setup}
\end{equation}

The inner edge agrees with the reference contour
$r_{\rm hit}/\mathcal M=6.2$ shown in Fig.~\ref{fig:inner_contour}. The
geometrical and radiative calculations therefore use the same inner disc
boundary.

Only the first equatorial intersection of each ray is retained. The images in
this section consequently represent the direct image of the disc. The
higher-order crossings associated with the additional transfer-function
branches in Sec.~\ref{sec:tf_results} are not included.


The frequency shift between the emitter and the observer is defined by
\begin{equation}
g\equiv\frac{\nu_{\rm obs}}{\nu_{\rm em}}
=
\frac{p_\mu u^\mu_{\rm obs}}{p_\mu u^\mu_{\rm em}},
\label{eq:redshift_factor}
\end{equation}
where $p_\mu$ is the photon four-momentum, $u^\mu_{\rm obs}$ is the observer
four-velocity, and $u^\mu_{\rm em}$ is the four-velocity of the emitting
matter.

For the static observer introduced in Sec.~\ref{sec:raytracing},
\begin{equation}
u^\mu_{\rm obs}
=
\frac{\delta^\mu_t}
{\sqrt{g_{tt}(r_{\rm obs},\theta_{\rm obs})}}.
\label{eq:observer_four_velocity}
\end{equation}

The emitting matter is assumed to move on a circular equatorial geodesic,
\begin{equation}
u^\mu_{\rm em}=u^t(1,0,0,\Omega).
\label{eq:emitter_four_velocity}
\end{equation}

The normalization condition $u_\mu u^\mu=1$ leads to 
\begin{equation}
u^t
=
\left(g_{tt}+g_{\phi\phi}\Omega^2\right)^{-1/2}_{\theta=\pi/2},
\label{eq:ut_emitter}
\end{equation}
while the circular-geodesic condition yields
\begin{equation}
\Omega^2
=
-\left.\frac{\partial_r g_{tt}}{\partial_r g_{\phi\phi}}\right|_{\theta=\pi/2}
=
\frac{M(1+q)\left(1-\dfrac{2M}{r}\right)^{2q+1}}
{r^2\left[r-M(2+q)\right]}.
\label{eq:Omega_qmetric}
\end{equation}

For $q=0$, Eq.~\eqref{eq:Omega_qmetric} reduces to the Schwarzschild result
$\Omega^2=M/r^3$.

Since $p_t$ and $p_\phi$ are conserved along each null geodesic,
\begin{equation}
p_\mu u^\mu_{\rm em}
=
u^t\left(p_t+\Omega p_\phi\right).
\label{eq:emitter_factor}
\end{equation}

For the meridional rays used in the nearly face-on calculation, $p_\phi=0$,
so the frequency shift contains no longitudinal rotational Doppler term. For
the inclined observer, $p_\phi$ is generally nonzero, and both gravitational
and Doppler contributions are present.

Liouville's theorem gives
\begin{equation}
\frac{I_\nu}{\nu^3}=\mathrm{const}
\label{eq:liouville_intensity}
\end{equation}
along a null geodesic. After integration over frequency, the observed
bolometric intensity acquires a factor $g^4$. If several disc intersections
are retained, the intensity can be written as
\begin{equation}
I_{\rm obs}(X,Y)
=
\sum_m g_m^4(X,Y)\,
I_{\rm em}\!\left[r_m(X,Y)\right],
\label{eq:intensity_schematic}
\end{equation}
where the sum runs over the retained intersections
\citep{Cunningham1975,Luminet1979}. In the present calculation only the first
crossing is included, so that
\begin{equation}
I_{\rm obs}^{(1)}(X,Y)
=
g_1^4(X,Y)\,
I_{\rm em}\!\left[r_{\rm hit}(X,Y)\right].
\label{eq:direct_image_intensity}
\end{equation}

Equation~\eqref{eq:direct_image_intensity} is evaluated on the
two-dimensional observer-screen grid used for the radiative calculation.


\begin{figure*}[t]
\centering
\includegraphics[width=0.98\textwidth]
{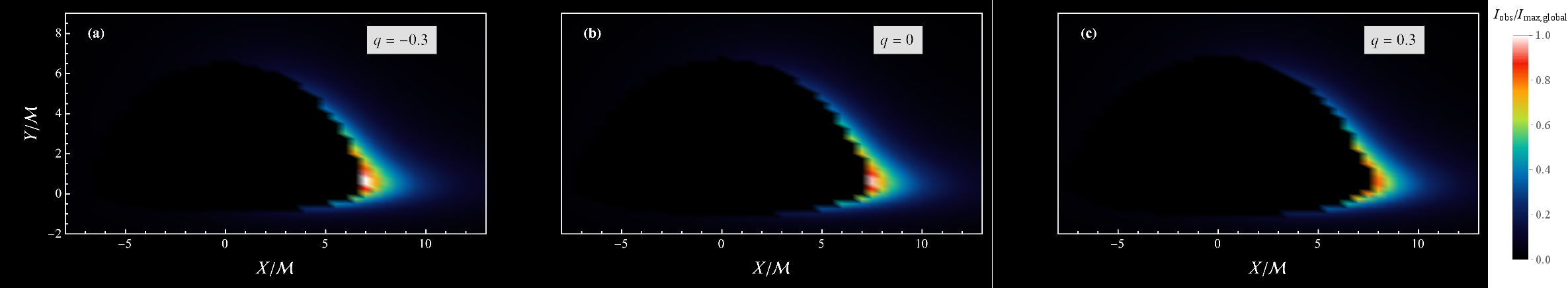}
\caption{Normalized observed-intensity maps of the direct disc image for
$q=-0.3$, $0$, and $0.3$ at fixed physical monopole mass. The observer is
located at $r_{\rm obs}/\mathcal M=1000$ and
$\theta_{\rm obs}=80^\circ$. The emitting disc has
$r_{\rm in}/\mathcal M=6.2$, $r_{\rm out}/\mathcal M=30$, and
$I_{\rm em}(r)\propto r^{-3}$. All panels are normalized by the same global
maximum intensity,
$I_{\max,\mathrm{global}}=\max_{q,X,Y} I_{\rm obs}(X,Y;q)$, so that relative
brightness differences between the three models are preserved. Only the
first equatorial intersection of each backward-traced ray is included.}
\label{fig:intensity_map}
\end{figure*}

Figure~\ref{fig:intensity_map} shows the direct images obtained for the three
representative values of $q$. The observer position, disc boundaries,
emissivity profile, and physical monopole mass are held fixed. All panels use
the common normalization
\begin{equation}
I_{\max,\mathrm{global}}
=
\max_{\substack{
q\in\{-0.3,\,0,\,0.3\}\\
(X,Y)\in\mathcal D}}
I_{\rm obs}(X,Y;q),
\label{eq:intensity_global_max}
\end{equation}
where $\mathcal D$ denotes the full computational screen. The plotted
quantity is
\begin{equation}
\widetilde I_{\rm obs}(X,Y;q)
=
\frac{I_{\rm obs}(X,Y;q)}{I_{\max,\mathrm{global}}}.
\label{eq:normalized_intensity}
\end{equation}

The common denominator preserves the relative brightness of the three
images. The field of view used in Fig.~\ref{fig:intensity_map} displays the
inner image region, whereas all integrated quantities reported below are
evaluated over the full computational screen.

The images have a similar overall shape, but the position and width of the
bright inner region vary with $q$. We examine this change by projecting the
intensity along the vertical screen direction:
\begin{equation}
\rho_q(X)
=
\int_{Y_{\min}}^{Y_{\max}} I_{\rm obs}(X,Y;q)\,dY.
\label{eq:projected_brightness}
\end{equation}

\begin{figure*}[t]
\centering
\includegraphics[width=0.98\textwidth]
{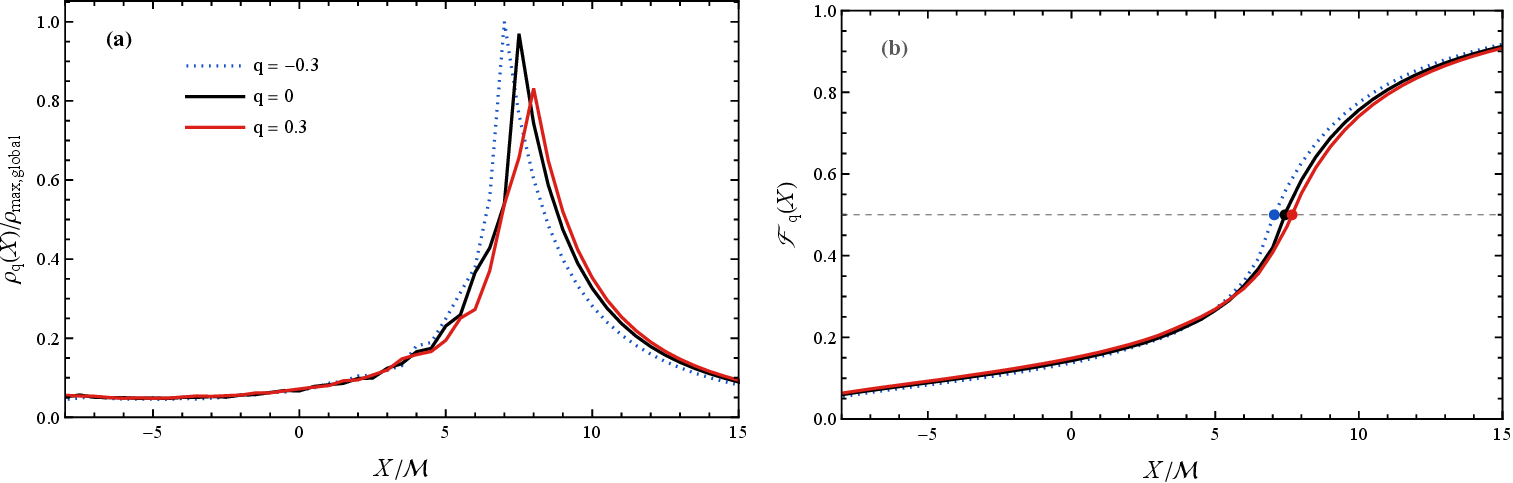}
\caption{Projected brightness profiles and cumulative-flux distributions of
the direct disc images for $q=-0.3$, $0$, and $0.3$. Panel (a) shows
$\rho_q(X)=\int I_{\rm obs}(X,Y;q)\,dY$, normalized by the common maximum
$\rho_{\max,\mathrm{global}}$ over all three models. Panel (b) shows the
corresponding cumulative-flux fractions $\mathcal F_q(X)$; the horizontal
dotted line marks $\mathcal F_q=0.5$. In both panels, the blue dotted, black
solid, and red solid curves correspond to $q=-0.3$, $0$, and $0.3$,
respectively. All profiles are evaluated over the full computational screen
at fixed physical monopole mass, with $\theta_{\rm obs}=80^\circ$,
$r_{\rm in}/\mathcal M=6.2$, $r_{\rm out}/\mathcal M=30$, and $p=3$. Only
the first equatorial intersection of each backward-traced ray is included.}
\label{fig:projected_profiles}
\end{figure*}

For Fig.~\ref{fig:projected_profiles}, the three profiles are divided by the
same maximum,
\begin{equation}
\rho_{\max,\mathrm{global}}
=
\max_{\substack{
q\in\{-0.3,\,0,\,0.3\}\\
X\in[X_{\min},X_{\max}]}}
\rho_q(X).
\label{eq:projected_brightness_global_max}
\end{equation}

This normalization retains their relative amplitudes. The cumulative-flux
fraction is defined as
\begin{equation}
\mathcal F_q(X)
=
\frac{\displaystyle\int_{X_{\min}}^{X}\rho_q(X')\,dX'}
{\displaystyle\int_{X_{\min}}^{X_{\max}}\rho_q(X')\,dX'}.
\label{eq:cumulative_flux}
\end{equation}

It satisfies $\mathcal F_q(X_{\min})=0$ and $\mathcal F_q(X_{\max})=1$. The median-flux position is determined by
\begin{equation}
\mathcal F_q(X_{50})=0.5.
\label{eq:median_flux_position}
\end{equation}

For $q=0$, the resolved third-crossing interval contains $b=3\sqrt{3}M$. The lower and upper quartiles are defined by $\mathcal F_q(X_{25})=0.25$ and
$\mathcal F_q(X_{75})=0.75$, respectively.


Figure~\ref{fig:projected_profiles}(a) shows a narrow main peak associated
with the bright inner part of the direct image. Its position and amplitude
change across the three models. The cumulative profiles in
Fig.~\ref{fig:projected_profiles}(b) provide a less peak-sensitive measure of
the horizontal displacement. 

The median-flux position increases monotonically:
\begin{equation}
\frac{X_{50}}{\mathcal M}
=
\begin{cases}
7.0554, & q=-0.3,\\
7.4289, & q=0,\\
7.6750, & q=0.3.
\end{cases}
\label{eq:x50_results}
\end{equation}

Relative to the Schwarzschild value, the shift is approximately $-5.0\%$
for $q=-0.3$ and $+3.3\%$ for $q=0.3$. The difference between the two
endpoint models is approximately $8.8\%$.

The central projected-flux interval also becomes wider as $q$ increases:

\begin{equation}
\frac{X_{75}-X_{25}}{\mathcal M}
=
\begin{cases}
4.9352, & q=-0.3,\\
5.2525, & q=0,\\
5.6499, & q=0.3.
\end{cases}
\label{eq:interquartile_width_results}
\end{equation}

This is an increase of approximately $14.5\%$ between $q=-0.3$ and
$q=0.3$. The individual value of $X_{25}$ is not monotonic. The projected
profiles should therefore not be interpreted as copies of one distribution
shifted rigidly along the screen: both their position and width change.


We next calculate a set of two-dimensional intensity-weighted diagnostics.
The total screen-integrated intensity is
\begin{equation}
F
=
\int_{\mathcal D} I_{\rm obs}(X,Y)\,dX\,dY,
\label{eq:radiative_flux}
\end{equation}
where $\mathcal D$ is the full computational screen. Since $I_0$ is
arbitrary, the flux is compared through ratios.

The intensity-weighted image centre is
\begin{align}
X_I
&=
\frac{1}{F}
\int_{\mathcal D}X I_{\rm obs}(X,Y)\,dX\,dY,
\nonumber\\
Y_I
&=
\frac{1}{F}
\int_{\mathcal D}Y I_{\rm obs}(X,Y)\,dX\,dY.
\label{eq:intensity_centroid}
\end{align}

The corresponding image widths are
\begin{align}
\sigma_X^2
&=
\frac{1}{F}\int_{\mathcal D}(X-X_I)^2 I_{\rm obs}(X,Y)\,dX\,dY,
\label{eq:sigma_x}\\
\sigma_Y^2
&=
\frac{1}{F}\int_{\mathcal D}(Y-Y_I)^2 I_{\rm obs}(X,Y)\,dX\,dY.
\label{eq:sigma_y}
\end{align}

We also calculate the intensity-weighted mean frequency-shift factor,
\begin{equation}
\langle g\rangle_I
=
\frac{\displaystyle\int_{\mathcal D}
g(X,Y)I_{\rm obs}(X,Y)\,dX\,dY}
{\displaystyle\int_{\mathcal D}I_{\rm obs}(X,Y)\,dX\,dY}.
\label{eq:mean_redshift}
\end{equation}

The integrals are evaluated as sums over the uniform screen grid. In
particular,
\begin{equation}
F
\simeq
\delta X\,\delta Y
\sum_{j\in\mathcal D} I_{{\rm obs},j},
\label{eq:discrete_flux}
\end{equation}
where $\delta X$ and $\delta Y$ are the grid spacings. Their common product
cancels from the normalized moments and from $\langle g\rangle_I$.


\begin{table}[t]
\centering
\small
\setlength{\tabcolsep}{2.6pt}
\caption{Intensity-weighted diagnostics of the direct images shown in
Fig.~\ref{fig:intensity_map}. The total observed flux is normalized to the
Schwarzschild value $F_0$. The centroid coordinates $(X_I,Y_I)$, image widths
$(\sigma_X,\sigma_Y)$, and intensity-weighted mean frequency-shift factor
$\langle g\rangle_I$ are evaluated from the unnormalized observed intensities
over the full computational screen. The common global normalization used in
Fig.~\ref{fig:intensity_map} is applied only for display.}
\label{tab:radiative_diagnostics}
\begin{tabular}{ccccccc}
\hline
$q$ & $F/F_0$ & $X_I/\mathcal M$ & $Y_I/\mathcal M$ &
$\sigma_X/\mathcal M$ & $\sigma_Y/\mathcal M$ & $\langle g\rangle_I$
\tabularnewline
\hline
$-0.3$ & $0.976$ & $6.632$ & $2.229$ & $7.467$ & $3.242$ & $1.109$
\tabularnewline
$0$ & $1.000$ & $6.780$ & $2.276$ & $7.708$ & $3.318$ & $1.103$
\tabularnewline
$0.3$ & $0.994$ & $6.845$ & $2.294$ & $7.905$ & $3.384$ & $1.097$
\tabularnewline
\hline
\end{tabular}
\end{table}

Table~\ref{tab:radiative_diagnostics} shows that the total direct-image flux
changes only weakly over the interval $-0.3\leq q\leq0.3$. The largest
departure from the Schwarzschild value is approximately $2.4\%$, occurring
for $q=-0.3$. At fixed physical monopole mass, the adopted emission model
therefore gives only a small change in the integrated flux.

The spatial diagnostics show a clearer trend. The horizontal centroid
increases from $X_I/\mathcal M=6.632$ to $6.845$, while the vertical centroid
changes from $2.229$ to $2.294$. Over the same interval,
$\sigma_X/\mathcal M$ increases from $7.467$ to $7.905$, and
$\sigma_Y/\mathcal M$ from $3.242$ to $3.384$. The displacement is larger in
the horizontal direction, although both widths increase.

The intensity-weighted mean frequency-shift factor decreases from
$\langle g\rangle_I=1.109$ at $q=-0.3$ to $1.097$ at $q=0.3$. This small
change includes the effects of the emitter motion, photon propagation, and
the redistribution of intensity over the screen.

The projected-profile and two-dimensional diagnostics describe different
features of the same intensity distribution. The centroid and standard
deviation include the extended image wings, whereas the median and
interquartile width depend mainly on the central part of the projected flux.
Both sets of diagnostics show a horizontal displacement and an increase in
image width as $q$ grows.

The radiative trends are also consistent with the geometrical results in
Sec.~\ref{sec:diagnostics}. The dimensions of the first-intersection contours
increase with $q$, and the intensity-weighted widths follow the same ordering.
The integrated flux, by contrast, remains close to its Schwarzschild value.
Within the present emission prescription, the main change is therefore a
redistribution of the direct-image brightness rather than a large variation
in its total value.

The numerical evaluation of the full-screen radiative diagnostics and the
independent checks based on the normalized intensity data are described in~\ref{app:radiative_diagnostics}.

The results of this section depend on the prescribed emissivity law and on
the restriction to the first equatorial intersection. A radiative treatment
of the second and third transfer-function branches would require successive
disc crossings to be retained and weighted separately. The images considered
here should therefore be read as a direct-image realization of the
first-intersection geometry, not as a complete model of an astrophysical
accretion flow.

\section{Discussion}
\label{sec:discussion}

The calculations presented in this work address three stages of the disc-imaging problem. The transfer functions identify successive equatorial crossings of null geodesics, the inclined first-intersection maps describe the two-dimensional disc-to-screen mapping, and the radiative calculation assigns an observed intensity to the direct image. This separation is important because the transfer functions and contour diagnostics are purely geometrical, whereas the intensity maps also depend on the emissivity profile and the motion of the emitting matter.

The strongest response to the deformation in the nearly face-on calculation is found in the third branch of the transfer-function. Its centre moves towards larger impact parameters as $q$ increases when the result is expressed in units of the metric parameter $M$. The analytic critical impact parameter associated with the unstable equatorial circular null orbit follows the same ordering. These two quantities are not identical: $b_{\rm cr}$ is a limiting photon-orbit scale, while $b_{3,c}$ is the midpoint of a finite interval of rays that reach the disc for the third time. Their similar behaviour nevertheless confirms that the third branch is controlled by photon trajectories close to the critical region.

{The distinction becomes particularly informative after normalization by the physical monopole mass. In that normalization, $b_{3,c}/\mathcal M$ decreases slightly from about $5.309$ to $5.156$ as $q$ increases from $-0.3$ to $0.3$, while $b_{\rm cr}/\mathcal M$ increases from about $4.906$ to $5.298$. Thus, the finite third-crossing branch and the limiting critical scale exhibit opposite residual trends once the dominant monopole rescaling is removed. This shows that $b_{3,c}$ is not merely a numerical proxy for $b_{\rm cr}$: it also contains information about the global disc-to-observer mapping required for a photon to complete a third equatorial crossing.}

The size of this displacement strongly depends on the mass normalization. For the $q$-metric, the physical monopole mass is
$\mathcal M=(1+q)M$. Comparing different values of $q$ at fixed $M$ therefore changes the monopole mass together with the quadrupolar deformation. Once the same results are expressed in units of $\mathcal M$, the variation of the third-branch position becomes considerably smaller. Thus, a substantial part of the displacement visible in $b_{3,c}/M$ is associated with the change in the monopole scale. The residual variation in $b_{3,c}/\mathcal M$ gives a more appropriate measure of the deformation effect at fixed physical mass.

{Equivalently, a comparison performed at fixed $M$ mixes two effects: the change of the monopole scale implied by $\mathcal M=(1+q)M$ and the genuine quadrupolar deformation of the null-geodesic map. Expressing the observables in units of $\mathcal M$, or comparing configurations at fixed $\mathcal M$, largely removes the first contribution and provides a cleaner diagnostic of the second.}

The finite width of the resolved third-crossing interval also changes with $q$. This width is determined by the set of numerically identified rays that produce a third disc crossing and should not be interpreted as the width of an observable photon ring. The branch center is therefore the most useful diagnostic for comparing the three geometries. Its convergence under grid refinement also makes it less sensitive to the precise numerical location of the interval boundaries.

The inclined first-intersection maps provide a separate test of the geometry. These maps contain neither emissivity nor frequency-shift weighting, and all contour comparisons are performed at fixed $\mathcal M$. Their differences consequently measure changes in the disc-to-screen mapping after the monopole rescaling has been removed.

The contour diagnostics show a clear radial dependence. The largest relative differences occur for the innermost reference contours, including the emitting-disc boundary at $r_{\rm hit}/\mathcal M=6.2$. The differences decrease at larger disc radii, where the photon trajectories sample a weaker-field region. This is consistent with the transfer-function results: the deformation is most evident for rays that pass close to the critical photon region.

Both the horizontal and vertical contour dimensions increase with $q$, although not by the same relative amount. The deformation therefore produces a mild anisotropic change rather than a uniform rescaling of the projected disc. The contour topology remains unchanged over the interval $-0.3\le q\le0.3$, and the reflection-symmetry diagnostics remain within the numerical accuracy of the calculation. These checks indicate that the measured contour displacements are features of the ray mapping rather than grid-induced asymmetries.

The radiative calculation shows how this geometrical change appears in a direct disc image. With the emissivity law, disc boundaries, observer inclination, and physical monopole mass held fixed, the total flux differs only slightly between the three models. The largest deviation from the Schwarzschild value is approximately $2.4\%$. By comparison, the spatial distribution of the intensity shows a clearer response.

In particular, the median-flux position changes from \\
$X_{50}/\mathcal{M} = 7.0554$ at $q=-0.3$ to $7.4289$ at $q=0$ and $7.6750$ at $q=0.3$. The difference between the two endpoint models is approximately $8.8\%$. At the same time, the interquartile width of the projected profile increases from
$(X_{75}-X_{25})/\mathcal M=4.9352$ to $5.6499$, corresponding to an increase of approximately $14.5\%$. Since $X_{25}$ itself is not monotonic, the profiles cannot be described as copies of a single distribution translated along the screen. The projected brightness is displaced and changes its width.

The two-dimensional moments support this interpretation. Across the same deformation interval, the horizontal intensity centroid increases from
$X_I/\mathcal M=6.6316$ to $6.8449$, while
$\sigma_X/\mathcal M$ increases from $7.4669$ to $7.9053$. The vertical centroid and width also increase, but their changes are smaller. The intensity-weighted mean frequency-shift factor decreases slightly, from $1.109$ to $1.097$. The observed differences therefore arise from the combined change in the geometrical mapping, circular emitter motion, and frequency-shift weighting. The present calculation does not isolate these contributions individually.

The projected-profile and two-dimensional diagnostics serve different purposes. The centroid and standard deviation use the full intensity distribution and are influenced by its extended wings. The median and interquartile width depend mainly on the central part of the projected flux and are less sensitive to the tails. Their consistent ordering shows that the horizontal displacement and broadening are not artifacts of a single choice of image statistic.

The geometrical and radiative results are compatible, but they should not be identified with one another. The contour maps determine where rays from given disc radii appear on the observer screen. The radiative image additionally weights these rays by $I_{\rm em}(r)$ and $g^4$. For the adopted prescription, the increase in contour dimensions with $q$ is accompanied by larger intensity-weighted image widths, while the integrated flux remains nearly constant. The main radiative effect is therefore a redistribution of the direct-image brightness rather than a uniform enhancement or suppression of the entire image.

{At fixed physical monopole mass, the quadrupolar deformation therefore manifests itself more clearly through spatial brightness diagnostics than through the integrated flux. In the present model, the endpoint change is about $8.8\%$ in the median-flux position and $14.5\%$ in the interquartile width, compared with a maximum flux difference of only about $2.4\%$ relative to Schwarzschild.}

This conclusion is specific to the model considered here. The emissivity is prescribed as a radial power law, the disc is geometrically thin, and the emitting matter follows circular equatorial geodesics. Only the first equatorial intersection is retained in the intensity calculation. The images consequently provide a controlled radiative realization of the first-intersection geometry, not a complete model of an astrophysical accretion flow.

Higher-order crossings may be particularly relevant close to the critical curve, where the second and third transfer-function branches are formed. Their radiative contributions cannot be inferred from the direct image alone and require successive disc intersections to be retained during ray tracing. A more complete calculation would also need to consider the optical properties and dynamical structure of the emitting material. Finite optical depth, disc thickness, magnetic fields, non-geodesic motion, and time-dependent emission may alter the relative importance of the image components.

The $q$-metric itself should likewise be understood as an exact static vacuum geometry used to study the effect of a quadrupolar deformation. For $q\ne0$, the surface $r=2M$ is singular rather than a regular event horizon. The present results therefore describe photon propagation and disc imaging in this exterior geometry and should not be interpreted as a model-independent prediction for black-hole observations.

{Accordingly, the quantities identified here are best regarded as geometrical and radiative imaging signatures, or diagnostics, of the quadrupolar exterior geometry. Establishing direct observational signatures would require a more complete emission model and a systematic comparison with instrumental observables.}

Despite these restrictions, the three parts of the calculation give a consistent picture. The near-critical transfer-function branch is the most sensitive one-dimensional geometrical diagnostic. The inclined contours show that the deformation is strongest in the inner projected disc and becomes weaker with increasing emission radius. After radiative weighting, the total direct-image flux changes little, but the brightness distribution shifts and broadens. Comparisons at fixed physical monopole mass are essential in all three cases, since otherwise the deformation effect is mixed with a change in the overall monopole scale.

\section{Conclusions}
\label{sec:conclusions}

We have studied thin-disc imaging in the static axisymmetric $q$-metric using nearly face-on transfer functions, inclined first-intersection maps, and a minimal radiative model. The calculations were organized so that the geometrical ray mapping could be examined separately from the emissivity and frequency-shift weighting used to construct the observed intensity.

The third transfer-function branch shows the clearest near-critical response to the deformation. Its centre follows the same qualitative ordering as the analytic critical impact parameter of the unstable equatorial null orbit. The comparison also demonstrates the importance of mass normalization. The displacement is pronounced in units of the metric parameter $M$, but is substantially reduced when the results are compared at fixed physical monopole mass,
$\mathcal M=(1+q)M$. The remaining shift represents the deformation effect after the changing monopole scale has been removed.

{An additional result emerges in this fixed-$\mathcal M$ comparison: $b_{3,c}/\mathcal M$ decreases mildly with increasing $q$, whereas the analytic critical scale $b_{\rm cr}/\mathcal M$ increases. The third-branch centre should therefore not be interpreted as a numerical estimate of the critical impact parameter. It is a distinct finite-crossing diagnostic that depends on the full propagation from the observer to the disc.}

The inclined first-intersection maps extend the analysis to the full observer screen. Their contour diagnostics show that the inner disc mapping is more sensitive to $q$ than the outer part. The horizontal and vertical contour dimensions increase with the deformation parameter, with a small change in aspect ratio, while the topology and expected reflection symmetry of the maps are preserved.

For the radiative illustration, we adopted a thin disc with
$r_{\rm in}/\mathcal M=6.2$,
$r_{\rm out}/\mathcal M=30$, and
$I_{\rm em}\propto r^{-3}$, viewed at an inclination of $80^\circ$. Only the first equatorial intersection was included. The total direct-image flux changes by no more than about $2.4\%$ relative to the Schwarzschild case over the range considered. The spatial intensity diagnostics show a larger effect. The median-flux position increases from
$X_{50}/\mathcal M=7.0554$ at $q=-0.3$ to $7.6750$ at $q=0.3$, an endpoint change of approximately $8.8\%$. Over the same interval, the interquartile width increases from $4.9352\mathcal M$ to $5.6499\mathcal M$, or by approximately $14.5\%$.

The intensity centroid and two-dimensional image widths follow the same general ordering. These results show that, within the adopted emission model, the quadrupolar deformation affects the spatial distribution of the direct-image brightness more clearly than its integrated flux. The change is not a simple translation: the projected intensity distribution also broadens.

{Thus, within the controlled direct-image model used here, the most sensitive radiative imprint of the deformation is the redistribution of brightness across the screen rather than a substantial modification of the total received flux.}

The radiative results remain dependent on the prescribed emissivity and include no higher-order image contributions. Retaining successive disc crossings and introducing a self-consistent emission model are therefore necessary before making observational predictions. Such extensions can be implemented within the same ray-tracing framework and would allow the geometrical second- and third-crossing branches identified here to be assigned separate radiative contributions.

The main result of the present analysis is that the deformation leaves a consistent, radius-dependent imprint on the disc-to-screen mapping, but its apparent magnitude depends strongly on whether the metric parameter or the physical monopole mass is held fixed. Once this normalization issue is controlled, the remaining effect is moderate: it is concentrated in near-critical trajectories and the inner projected disc, and appears radiatively as a displacement and broadening of the direct-image brightness distribution.

\section*{Acknowledgements}
This work was partially supported by the Ministry of Science and Higher Education of the Republic of Kazakhstan, Grant No.{AP23489541}.
The work of HQ was supported by PAPIIT-DGAPA-UNAM, grant No. 108225, and CONAHCYT,  grant No. CBF-2025-I-253.
\section*{Data Availability}
 The data supporting the findings of this study are available from the corresponding author upon reasonable request.
\section*{Statements and Declarations}

\subsection*{Competing interests}
The authors declare that they have no competing interests.
\appendix
\section{Numerical implementation and consistency checks}
\label{app:numerics}

This appendix gives the numerical details needed to reproduce the ray-tracing
diagnostics discussed in the main text. The nearly face-on transfer functions
are first calculated in units of the metric parameter $M$. The inclined
first-intersection maps and the radiative images are compared at fixed
physical monopole mass,
$\mathcal M=(1+q)M$. For these calculations, the metric parameter is set
separately for each deformation according to
$M=\mathcal M/(1+q)$.

All photon trajectories are integrated as null geodesics of the fixed
$q$-metric background with signature $+---$, using the Hamiltonian system
given in Eqs.~\eqref{eq:Hamiltonian} and
\eqref{eq:Hamilton_equations}. The procedures described below do not introduce
additional physical assumptions. Their purpose is to specify how the ordered
disc crossings, the inclined contours, and the radiative diagnostics are
extracted from the numerical ray families.

\subsection{Integration domain and stopping criteria}
\label{app:integration_domain}

The one-dimensional transfer functions are calculated by tracing photons
backwards from the nearly face-on observer position specified in
Eq.~\eqref{eq:observer_faceon}. The integration is terminated when the ray
reaches the inner numerical cutoff
\begin{equation}
r\leq r_{\rm cut}^{\rm in},
\qquad
r_{\rm cut}^{\rm in}
=
\left(2+10^{-5}\right)M,
\label{eq:rin_cut_app}
\end{equation}
escapes through the outer boundary
\begin{equation}
r\geq r_{\rm max},
\qquad
r_{\rm max}=2000M,
\label{eq:rmax_app}
\end{equation}

or reaches the maximum affine parameter
\begin{equation}
0\leq\lambda\leq\lambda_{\rm max},
\qquad
\lambda_{\rm max}=5000.
\label{eq:lambda_range_app}
\end{equation}

The inner cutoff lies immediately outside the singular surface $r=2M$.
It is a numerical boundary and should not be confused with the inner edge of
the emitting disc. The transfer functions are defined by the recorded
equatorial crossings and do not use
$r_{\rm cut}^{\rm in}$ as a disc boundary. For all values of $q$
considered here, the equatorial circular photon orbit lies outside this
cutoff.

The geodesic equations are integrated with an adaptive solver and
${\tt MaxSteps}\to200000$. Both the accuracy goal and the precision goal
are set to $10$ in the production transfer-function runs. The conserved
momenta $p_t$ and $p_\phi$ are determined by the photon momentum in the
observer tetrad.

For each launched ray, the residual of the null constraint
\begin{equation}
H=\frac{1}{2}g^{\mu\nu}p_\mu p_\nu=0
\label{eq:hamiltonian_check_app}
\end{equation}
is evaluated at the observer. This provides a check on the tetrad-to-coordinate
transformation and the initial momentum assignment. It is particularly useful
for rays near the critical region, from which the higher-order
transfer-function branches are extracted.

\subsection{Disc-crossing detection and branch extraction}
\label{app:crossing_detection}

An equatorial crossing is detected when
\begin{equation}
\cos\theta(\lambda)=0
\label{eq:crossing_condition_app}
\end{equation}
and the corresponding radius satisfies
$r(\lambda)>r_{\rm cut}^{\rm in}$. The crossing location is refined by a
one-dimensional root search along the integrated trajectory. The detected
crossings are then ordered by increasing affine parameter along the backward
ray.

Numerical event detection can occasionally record the same crossing more than
once. Two events separated by less than
\begin{equation}
\Delta\lambda_{\rm dup}=10^{-5}
\label{eq:duplicate_lambda_app}
\end{equation}
are therefore treated as duplicate detections, and only one of them is
retained.

For a fixed $q$, the ordered crossing radii define the transfer-function
branches
$r_1(b;q),r_2(b;q),r_3(b;q),\ldots$, as introduced in
Eq.~\eqref{eq:crossing_sequence}. The first branch covers a broad interval in
the screen impact parameter and is already resolved by the initial scan. The
second and third branches occupy progressively narrower intervals close to the
strong-lensing region. Their extraction consequently requires a refined
sampling of $b$.

The scan is carried out in two stages. A coarse calculation first locates the
near-critical interval. The impact-parameter sampling is then refined over the
region containing the higher-order branches. The production step used for the
third-branch diagnostics is
\begin{equation}
\Delta b=5\times10^{-4}M.
\label{eq:db_sampling_app}
\end{equation}

The numerical centre $b_{3,c}$, defined in
Eq.~\eqref{eq:b3c_definition}, is an operational diagnostic of the resolved
third-crossing interval. It is not an exact geodesic invariant and is not
identified with the analytic critical impact parameter $b_{\rm cr}$.
The comparison between these quantities is used only to determine whether the
numerically resolved branch follows the expected near-critical photon scale.

\subsection{Resolution of the third-crossing branch}
\label{app:resolution_checks}

The principal numerical sensitivity of the one-dimensional transfer functions
comes from the finite sampling of the screen impact parameter. We therefore
treat the third branch as a resolved interval,
\begin{equation}
\left[b_{3,\min},b_{3,\max}\right],
\label{eq:b3_interval_app}
\end{equation}
rather than assigning exact numerical meaning to its endpoints. The centre and
width of this interval are
\begin{equation}
b_{3,c}
=\frac{1}{2}
\left(b_{3,\min}+b_{3,\max}\right)
\label{eq:b3c_app}
\end{equation}
and
\begin{equation}
\Delta b_3
=
b_{3,\max}-b_{3,\min},
\label{eq:db3_app}
\end{equation}
respectively. Changing the sampling step may move either endpoint by an amount
comparable to the grid spacing. The persistence of the branch and the
displacement of its centre are therefore more reliable than the last reported
digits of $b_{3,\min}$ and $b_{3,\max}$.

For a repeated calculation with step $\Delta b'$, the fractional change of
the branch centre is measured by
\begin{equation}
\epsilon_{3,c}(q;\Delta b')
=\left|
\frac{
b_{3,c}(q;\Delta b')
-
b_{3,c}(q;\Delta b)
}{
b_{3,c}(q;\Delta b)
}
\right|,
\label{eq:b3c_convergence_measure}
\end{equation}
where $\Delta b=5\times10^{-4}M$ is the production value. This convergence
measure is appropriate for a finite-interval midpoint. The branch centre is
regarded as resolved when its change under grid refinement is much smaller
than its total displacement across the deformation interval considered in the
main text.

The Schwarzschild case gives an independent reference. For $q=0$, the
analytic critical impact parameter is

\begin{equation}
b_{\rm cr}=3\sqrt{3}M,
\label{eq:schw_bcrit_app}
\end{equation}
as stated in Eq.~\eqref{eq:bcrit_schw}. This value lies within the resolved
third-crossing interval, as shown in Eq.~\eqref{eq:schw_validation}. The
crossing-detection procedure therefore recovers the expected near-critical
region in the Schwarzschild limit.

The main parameters of the production transfer-function calculation are
listed in Table~\ref{tab:numerical_setup}.

\begin{table}[t]
\centering
\caption{Numerical parameters used in the production calculation of the
nearly face-on transfer functions.}
\label{tab:numerical_setup}
\begin{tabular}{cc}
\hline
Quantity & Value
\tabularnewline
\hline
$r_{\rm obs}/M$ & $1000$
\tabularnewline
$\theta_{\rm obs}$ & $10^{-4}$
\tabularnewline
$r_{\rm cut}^{\rm in}/M$ & $2+10^{-5}$
\tabularnewline
$r_{\rm max}/M$ & $2000$
\tabularnewline
$\lambda_{\rm max}$ & $5000$
\tabularnewline
Near-critical $\Delta b/M$ & $5\times10^{-4}$
\tabularnewline
Accuracy goal & $10$
\tabularnewline
Precision goal & $10$
\tabularnewline
\hline
\end{tabular}
\end{table}

\begin{table}
\centering
\caption{Resolution check for the centre of the numerical
third-crossing branch. The production sampling used in the main text
is $\Delta b/M=5\times10^{-4}$. The fractional differences
$\epsilon_{3,c}$ are evaluated from the unrounded numerical values
and are measured relative to the production result.}
\label{tab:b3_convergence}
\begin{tabular}{cccc}
\hline
$q$ &
$\Delta b/M$ &
$b_{3,c}/M$ &
$\epsilon_{3,c}$
\\
\hline

$-0.3$ &
$10^{-3}$ &
3.7163 &
0
\\

$-0.3$ &
$5\times10^{-4}$ &
3.7163 &
0
\\

$-0.3$ &
$2.5\times10^{-4}$ &
3.7162 &
$3.36\times10^{-5}$
\\

$0$ &
$10^{-3}$ &
5.2027 &
0
\\

$0$ &
$5\times10^{-4}$ &
5.2027 &
0
\\

$0$ &
$2.5\times10^{-4}$ &
5.2027 &
0
\\

$0.3$ &
$10^{-3}$ &
6.7023 &
$3.73\times10^{-5}$
\\

$0.3$ &
$5\times10^{-4}$ &
6.7025 &
0
\\

$0.3$ &
$2.5\times10^{-4}$ &
6.7025 &
0
\\

\hline
\end{tabular}
\end{table}

Table~\ref{tab:b3_convergence} shows that refinement of the near-critical
impact-parameter sampling changes $b_{3,c}$ by much less than its total
displacement across the interval $-0.3\leq q\leq0.3 $. The monotonic
ordering of the third branch is therefore stable under the tested changes of
resolution.

\subsection{Inclined first-intersection maps}
\label{app:inclined_maps}

The inclined calculations use the same Hamiltonian equations and
equatorial-crossing condition as the one-dimensional transfer functions.
They differ in the observer orientation, the two-dimensional screen sampling,
and the mass normalization. The observer is placed at

\begin{equation}
\frac{r_{\rm obs}}{\mathcal M}=1000,
\qquad
\theta_{\rm obs}=80^\circ.
\label{eq:inclined_observer_app}
\end{equation}

Here $\theta_{\rm obs}$ is measured from the symmetry axis and is equal to
the disc inclination used in the main text.

The computational screen covers
\begin{equation}
-40\leq\frac{X}{\mathcal M}\leq40,
\qquad
-10\leq\frac{Y}{\mathcal M}\leq20.
\label{eq:screen_domain_app}
\end{equation}

The full screen is sampled with a $161\times121$ grid of independently
traced rays for each value of $q$. Reflection symmetry is not imposed in
constructing the production maps. The expected symmetry under
$X\rightarrow-X$ is instead used as an independent check of the numerical
calculation.

Each screen point is assigned the radius of the first accepted equatorial
intersection,
\begin{equation}
r_{\rm hit}=r_{\rm hit}(X,Y;q).
\label{eq:first_hit_map_app}
\end{equation}

The reference contours in Sec.~\ref{sec:diagnostics} are extracted as the
level sets
\begin{equation}
\frac{r_{\rm hit}(X,Y;q)}{\mathcal M}
=
\frac{r_0}{\mathcal M},
\qquad
\frac{r_0}{\mathcal M}=6,10,20,30.
\label{eq:contour_levels_app}
\end{equation}

The additional level
\begin{equation}
\frac{r_{\rm hit}(X,Y;q)}{\mathcal M}=6.2
\label{eq:inner_contour_level_app}
\end{equation}
is extracted from the same maps. It coincides with the inner emitting-disc
boundary used in Sec.~\ref{sec:emission} and provides a direct geometrical
reference for the radiative image. Neither the reference contours nor the
$6.2\mathcal M$ contour use emissivity or frequency-shift weighting.

The contour diagnostics in
Table~\ref{tab:contour_diagnostics} are calculated from the extracted level
sets. The quantities
$\Delta X$, $\Delta Y$, $\mathcal A_{\rm c}$,
$X_{\rm c}^{\rm geom}$, and $Y_{\rm c}^{\rm geom}$ are therefore
properties of the numerical disc-to-screen map and are not
intensity-weighted observables.

The full-screen calculation also provides a direct symmetry test. The largest
normalized horizontal-centre residual is
\begin{equation}
\max\left(
\frac{\left|X_{\rm c}^{\rm geom}\right|}{\Delta X}
\right)
=7.3\times10^{-5},
\label{eq:x_symmetry_residual_app}
\end{equation}
and the contour widths obtained from the full-screen and reflected
constructions differ by less than $2\times10^{-4}$. These residuals are
small compared with the deformation-dependent differences reported in
Table~\ref{tab:contour_diagnostics}. The observed ordering of the contours is
therefore not caused by a measurable left--right bias of the numerical grid.

Because the contours are extracted from a finite screen sampling, their
reported dimensions should be understood as numerical diagnostics of the
ray-tracing map rather than as precision observables. Their role is to
quantify the systematic change of the first-intersection geometry with $q$
at fixed physical monopole mass.

\subsection{Radiative maps and integrated diagnostics}
\label{app:radiative_diagnostics}

The direct-image calculation uses the first equatorial intersection stored for
each accepted ray. A nonzero intensity is assigned only when the crossing
radius satisfies
\begin{equation}
6.2
\leq
\frac{r_{\rm hit}}{\mathcal M}
\leq
30.
\label{eq:radiative_acceptance_app}
\end{equation}

For these rays, the observed bolometric intensity is evaluated from
Eq.~\eqref{eq:direct_image_intensity}. Rays that miss the emitting interval,
reach the inner numerical cutoff, or leave the integration domain without an
accepted disc intersection are assigned zero intensity.

The radiative maps are evaluated on the full computational screen
\begin{equation}
-40\leq\frac{X}{\mathcal M}\leq40,
\qquad
-10\leq\frac{Y}{\mathcal M}\leq20,
\label{eq:radiative_screen_domain_app}
\end{equation}
using a uniform $161\times121$ grid. The smaller field of view used in
Fig.~\ref{fig:intensity_map} is applied only when the figure is displayed.
It does not enter the flux, centroid, width, frequency-shift, or projected
profile calculations.

Let $I_{ij}(q)$ denote the unnormalized intensity at a point of the uniform
screen grid. The discrete form of the total flux is
\begin{equation}
F(q)
\simeq
\delta X \,\delta Y
\sum_{i,j}I_{ij}(q),
\label{eq:discrete_flux_app}
\end{equation}
where the sum covers the full computational screen. The same full-screen sum
is used for the intensity centroid and second moments. Since the grid is
uniform, the common area factor $\delta X\,\delta Y$ cancels from the
normalized moments, the flux ratios, and the intensity-weighted mean
frequency-shift factor.

The displayed images use the common normalization
\begin{equation}
I_{\max,\mathrm{global}}
= \max_{\substack{
q\in\{-0.3,0,0.3\}\
i,j}}
I_{ij}(q).
\label{eq:global_normalization_app}
\end{equation}

Thus, every panel in Fig.~\ref{fig:intensity_map} is divided by the same
number. This preserves the relative peak intensities of the three models.
It differs from a panel-by-panel normalization, which would set the maximum
of each image to unity and remove this information.

The integrated diagnostics in
Table~\ref{tab:radiative_diagnostics} are calculated from the unnormalized
intensity arrays. As a numerical check, the flux ratios, centroids, and image
widths were independently recomputed from the globally normalized arrays.
Because the same constant multiplies every pixel in all three images, it
cancels from the normalized moments and flux ratios. The recomputed values
agree with the stored diagnostics to machine precision.

The projected profile is evaluated as a discrete sum along the vertical
screen direction,
\begin{equation}
\rho_q(X_i)
\simeq
\delta Y
\sum_j I_{ij}(q).
\label{eq:projected_profile_discrete_app}
\end{equation}

Its cumulative distribution is then constructed from
\begin{equation}
\mathcal F_q(X_k)
=\frac{
\displaystyle
\sum_{i\leq k}\rho_q(X_i) \,\delta X
}{
\displaystyle
\sum_i\rho_q(X_i) \delta X
}.
\label{eq:cumulative_flux_discrete_app}
\end{equation}

The positions $X_{25}$, $X_{50}$, and $X_{75}$ are obtained by
interpolating the cumulative distribution at the levels $0.25$, $0.50$,
and $0.75$, respectively. The interquartile width is then calculated as
\begin{equation}
W_{50}=X_{75}-X_{25}.
\label{eq:interquartile_width_app}
\end{equation}

The projected profiles and cumulative distributions shown in
Fig.~\ref{fig:projected_profiles} are therefore derived from the same
full-screen intensity data as the two-dimensional diagnostics in
Table~\ref{tab:radiative_diagnostics}. The projected and two-dimensional
quantities are not separate image calculations; they are complementary
summaries of the same direct-image data.

\bibliographystyle{spphys}
\bibliography{references}

\label{lastpage}
\end{document}